\documentclass[
nofootinbib,
noeprint,
amsmath,amssymb,
aps,
nobibnotes,
twocolumn,
floatfix,
]{revtex4-2}

\usepackage[utf8]{inputenc}
\usepackage{graphicx}
\usepackage{amsmath}
\usepackage{amssymb}
\usepackage{bm}
\usepackage{txfonts}
\usepackage{multirow}
\usepackage{mathptmx} 
\usepackage{siunitx}
\usepackage{xargs}
\usepackage{printlen}
\usepackage{booktabs}
\usepackage[pdftex,dvipsnames]{xcolor}
\usepackage[colorlinks=true, urlcolor=blue, linkcolor=blue, citecolor=blue]{hyperref}
\usepackage{xcolor}
\usepackage{braket}
\usepackage{soul}
\usepackage{lipsum}
\usepackage{mathrsfs}
\usepackage{verbatim}
\begin{document}

\begin{abstract}
Quantum simulations of lattice gauge theories (LGTs) with both dynamical matter and gauge fields provide a promising approach to studying strongly coupled problems beyond classical computational reach. Yet, implementing gauge-invariant encodings and real-time evolution remains experimentally challenging. Here, we demonstrate a resource-efficient encoding of a $\mathbb{Z}_2$ LGT using a hybrid qubit-oscillator trapped-ion quantum device, where qubits represent gauge fields and vibrational modes naturally encode bosonic matter fields. This architecture utilises synthetic dimensions to construct higher-dimensional lattice geometries and combines digital and analogue techniques to prepare initial states, realise gauge-invariant real-time evolution, and measure the relevant observables. We experimentally probe dynamics obeying Gauss’s law in a $\mathbb{Z}_2$ link and extend this to a loop geometry, marking the first steps towards higher-dimensional LGTs. 
In this quasi-2D setup, we observe Aharonov-Bohm interference for the first time with dynamical gauge fields encoding magnetic flux, demonstrating the interplay between charge and flux.
Our results chart a promising path for scalable quantum simulations of bosonic gauge theories and outline a roadmap for realising exotic LGTs in higher dimensions.
\end{abstract}
\title{Real-Time Observation of Aharonov-Bohm Interference in a $\mathbb{Z}_2$ Lattice Gauge Theory on a Hybrid Qubit-Oscillator Quantum Computer}
\date{\today}
\author{S. Saner$^1$, O. B\u{a}z\u{a}van$^1$, D. J. Webb$^1$, G. Araneda$^1$, C. J. Ballance$^1$, R. Srinivas$^1$, D. M. Lucas$^1$, A. Berm\'udez$^{1,2}$\\
\normalsize{$^1$Department of Physics, University of Oxford, Clarendon Laboratory, \\
Parks Road, Oxford OX1 3PU, United Kingdom\\
$^2$Instituto de F\'isica Teorica, UAM-CSIC, Universidad Autónoma de Madrid,\\ Cantoblanco, 28049 Madrid, Spain
\\Email: sebastian.saner@physics.ox.ac.uk}}

\maketitle

Gauge theories provide a unifying framework for understanding the
interactions of matter and fields in nature, ranging from high-energy physics~\cite{doi:10.1073/pnas.93.25.14256} to condensed matter~\cite{doi:10.1098/rsta.2015.0248}.
Matter fields interact 
through gauge fields 
according to laws that are invariant under specific local symmetries.
In the Standard Model of particle physics~\cite{peskin1995intro}, for example, the electromagnetic, weak, and strong forces are mediated by gauge fields in three spatial and one temporal dimension ($3+1$D)~\cite{PhysRev.96.191}. 
These gauge fields arise from local symmetries described by the gauge groups ${\rm U}(1)$, ${\rm SU}(2)$, and ${\rm SU}(3)$, respectively. 
While gauge theories 
underpin particle physics, they also arise as effective descriptions of strongly-correlated condensed-matter systems.
However, solving them in physically relevant regimes is often challenging. In particle physics, phenomena like deep inelastic scattering and dense nuclear matter, involving real-time dynamics and finite-density, respectively, elude standard Monte Carlo approaches~\cite{PhysRevLett.94.170201,nagata2022finite}. Similarly, in condensed matter, strongly-correlated systems and topological order require non-perturbative approaches.

Lattice gauge theories (LGTs), which discretise models onto lattices with gauge fields on links and matter at vertices provide a powerful framework for addressing these challenges~\cite{PhysRevD.10.2445,kogut1979an, PhysRevD.11.395}. Among LGTs, those with discrete $\mathbb{Z}_2$ groups~\cite{wegner1971duality,PhysRevD.19.3682} are particularly significant in condensed matter physics, as they describe frustrated antiferromagnets~\cite{PhysRevLett.66.1773}, high-temperature superconductors~\cite{PhysRevB.37.580,PhysRevB.62.7850}, and quantum spin-liquid phases~\cite{kitaev2006anyons}.
In the context of quantum information processing, $\mathbb{Z}_2$ LGTs~\cite{wegner1971duality,PhysRevD.19.3682} are relevant to topological quantum error correction~\cite{kitaev2003fault, nielsen00}.

The richness of phenomena associated with $\mathbb{Z}_2$ LGTs 
grows with the system's dimensionality.
For the case of $1+1$D, we can observe gauge-invariant tunnelling along with signatures of confinement, where the propagation of matter fields starts to be restricted~\cite{PhysRevLett.124.120503,PhysRevB.109.245110}. In $2+1$D dimensions, the lattice can be pierced by magnetic flux, resulting in quasi-particle excitations called visons~\cite{PhysRevB.62.7850}. In these electrically and magnetically-dominated phases, the real-time dynamics of the gauge fields are intertwined with confined or deconfined matter fields. In $3+1$D, the study of flux sources, i.e., discrete analogues of magnetic monopoles, becomes possible. Understanding these strongly-correlated, real-time dynamics is
a classically intractable challenge, 
and the use of quantum simulators could offer a solution~\cite{feynman1982simulating,zohar2016quantum,Banuls2020simulating,dalmonte2016lattice,Aidelsburger2021cold,PRXQuantum.4.027001,PRXQuantum.5.037001,Halimeh:2025lid}.

In recent years, experimental demonstrations of quantum simulations of real-time dynamics in $\mathbb{Z}_2$ LGTs have mainly focused on $1+1$D
geometries, where visons and topological effects are absent. These experiments have implemented
analogue Floquet schemes with cold atoms in a Hubbard double-well link~\cite{Schweizer2019floquet}, 
or digital gate-based schemes with superconducting qubits using linear connectivity~\cite{mildenberger2025confinement,alexandrou2025realizingstringbreakingdynamics}. Both analogue and digital approaches have also employed effective formulations of $2+1$D systems
by enforcing Gauss’s law to eliminate
matter degrees of freedom~\cite{cochran2025visualizing,de2024observationstringbreakingdynamicsquantum,farrell2025digitalquantumsimulationsscattering,luo2025quantumsimulationbubblenucleation}. 
These approaches, while offering insight, cannot experimentally observe the interplay between dynamical gauge fields and matter fields, which gives rise to the attachment of electric field lines to charges via Gauss’s law, as well as flux-charge interactions in higher-dimensional systems.

Here, to address these gaps, we demonstrate a quantum simulation of a $\mathbb{Z}_2$ LGT using a trapped-ion processor with a hybrid qubit-oscillator architecture, where qubits encode gauge fields via electronic levels and oscillators represent matter fields via vibrational modes~\cite{bazavan2024synthetic, davoudi2021toward,crane2024hybridoscillatorqubitquantumprocessors}. Our approach is hybrid in two ways. First, it combines discrete and continuous variables to efficiently simulate the target LGT. Second, it integrates digital operations (single- and two-qubit, and qubit-oscillator gates), for initialisation, error suppression, and measurement, with analogue evolution under an engineered $\mathbb{Z}_2$-invariant Hamiltonian (Fig.~\ref{fig:scheme_link}(a,b)).

The hybrid architecture enables the realisation of lattice gauge geometries by introducing synthetic spatial dimensions via qubit-oscillator couplings. We demonstrate correlated real-time dynamics of a matter charge and an attached electric-field line, consistent with Gauss’s law in a minimal $1+1$D $\mathbb{Z}_2$ link encoded in a single trapped ion. 
Extending this approach to a two-ion crystal, we add a second $\mathbb{Z}_2$ link to form a loop, enabling fundamental $2+1$D gauge dynamics. In this configuration, we prepare entangled gauge fields representing a gauge-invariant $\mathbb{Z}_2$ flux piercing the loop and demonstrate that charge dynamics are inhibited by destructive interference depending on the flux state (i.e., the presence of a vison); this provides a direct observation of the $\mathbb{Z}_2$ Aharonov-Bohm effect.

\section*{lattice gauge theory encoding}
\begin{figure}
    \centering
    \includegraphics[]{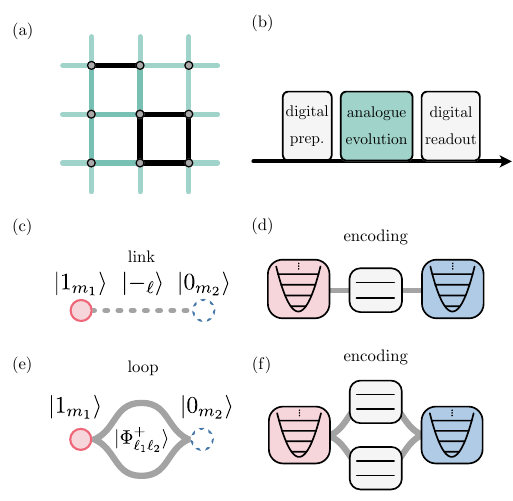}
    \caption{Quantum simulation of a \(\mathbb{Z}_2\) lattice gauge field theory using the hybrid encoding scheme.  
(a) The gauge fields reside on the links connecting the matter sites, which are located at the nodes. The fundamental building blocks, the gauge-matter link and the loop, are highlighted in black.  
(b) The quantum simulation pulse sequence consists of a digital preparation of the hybrid system used to encode the LGT followed by analogue evolution under the gauge-invariant Hamiltonian $\hat{H}_{\mathbb{Z}_2}$ and concludes with digital readout.
(c) Illustration of a single gauge-matter link in state $\ket{1_{m_1}, -_\ell, 0_{m_2}}$ and (d) the corresponding encoding into the hybrid qubit-oscillator system. The matter sites are represented by two motional modes of a harmonically confined trapped ion, while the gauge field is encoded in two levels of its internal electronic structure.  
(e) Illustration of a $\mathbb{Z}_2$ loop in state $\ket{1_{m_1}, \Phi_{\ell_1\ell_2}^+, 0_{m_2}}$ with the gauge fields prepared in the entangled state $\ket{\Phi_{\ell_1\ell_2}^+}$ and (f) the corresponding extension of the hybrid encoding with an additional gauge field.
}
    \label{fig:scheme_link}
\end{figure}

We take a bottom-up approach to LGTs, starting with the simplest element: a $\mathbb{Z}_2$ link, where two matter sites are coupled via a gauge field (Fig.~\ref{fig:scheme_link}(c)). From this basic unit, we progressively build more complex lattice geometries link by link.
Central to our approach is the encoding of the $\mathbb{Z}_2$ link within a qubit-oscillator quantum system (Fig.~\ref{fig:scheme_link}(d)). Here, the gauge field ($\ell$) is represented by a qubit, with its two states $\ket{\uparrow_{\ell}}$ and $\ket{\downarrow_{\ell}}$, acted on by 
Pauli matrices $\hat{\sigma}_{\ell}^{x,y,z}$. The matter sites $m_i$ with $i\in \{1, 2\}$, are encoded in harmonic oscillators,
with the bosonic particle creation ($\hat{a}_{m_i}^{{\dagger}}$) and annihilation ($\hat{a}_{m_i}^{\phantom{\dagger}}$) operators. This encoding enables the engineering of synthetic dimensions while naturally incorporating bosonic matter supporting multiple excitations on each site, extending the proposal in Ref.~\cite{bazavan2024synthetic}.

We consider an LGT that satisfies a local $\mathbb{Z}_2$ gauge symmetry, which in the single-link case is generated by $\hat{G}_{m_i} = \hat{\sigma}^x_\ell \hat{P}_{m_i}$ at each site and at all times, where $\hat{P}_{m_i} = \exp({\rm i} \pi \hat{a}_{m_i}^\dagger \hat{a}_{m_i}^{\phantom{\dagger}})$ is the matter parity. Physical states obey Gauss's law and are therefore eigenstates of $\hat{G}_{m_i}$, constraining
the gauge field to the two eigenstates of $\hat{\sigma}_{\ell}^x$; $ \ket{+_{\ell}} $ and $ \ket{-_{\ell}}$ corresponding to the presence ($+$) or absence ($-$) of an electric-field line connecting the two matter sites. The dynamical
charges are likewise constrained to eigenstates of $\hat{P}_{m_i} = 1 - 2 \hat{Q}_{m_i}$. Here, \( \hat{Q}_{m_i} \) is the bosonic \(\mathbb{Z}_2\) charge operator at site \( m_i \), with eigenvalues \( q_{m_i} \in \{0,1\} \). States with an even (odd) number of bosons $\ket{2k_{m_i}}(\ket{(2k+1)_{m_i}})$ have charge \( q_{m_i} = 0 \) (\(1\)).

When a matter particle is charged under a gauge group, it acquires an Aharonov-Bohm-type phase as it moves between lattice sites, determined by the line integral of the corresponding gauge field~\cite{PhysRevLett.3.296}. In the $\mathbb{Z}_2$ case, this leads to a real-time evolution where matter charges and gauge fields are coupled through the Hamiltonian:
\begin{equation}\label{eq:z2_gauge_link}
\hat{H}_{\mathbb{Z}_2}^{\rm link} =  \left(J\hat{a}_{m_1}^\dagger \hat{\sigma}^z_{\ell} \hat{a}_{m_2}^{\phantom{2}} + \text{H.c.}\right) + h \hat{\sigma}^x_{\ell}.
\end{equation} 
The first term describes matter tunnelling of strength $J$  between the two sites mediated by the 
gauge field. In the electric-field basis, each tunnelling event  flips the gauge field configuration ($ \ket{-_\ell} \leftrightarrow \ket{+_\ell} $).
The second term represents the electric-field energy, setting the cost $h$ of creating or destroying an electric field line between the matter sites.
Importantly, the evolution preserves the local gauge symmetry as $\hat{H}_{\mathbb{Z}_2}^{\rm link}$ commutes with the symmetry operators $[\hat{H}_{\mathbb{Z}_2}^{\rm link}, \hat{G}_{m_1}] = [\hat{H}_{\mathbb{Z}_2}^{\rm link}, \hat{G}_{m_2}] = 0$ (see Supplement~\cite{supplementary}). 
For an initial state \( \ket{\psi_0} \) that satisfies the symmetry constraint \( \hat{G}_{m_i} \ket{\psi_0} = g_{m_i} \ket{\psi_0} \), the system’s dynamics remain confined to the subspace defined by the eigenvalues \( (g_{m_1}, g_{m_2}) \), where \( g_{m_i} \in \{-1, 1\} \), as \( \hat{G}_{m_i}^2 = \mathbb{I} \). The values $g_{m_i} = -(-1)^{q_{m_i}^{\rm bg}}$ determine the background charges $q_{m_i}^{\rm bg}$, which are static. Electric field lines stretch between background and dynamic charges.


First, we focus on the $(g_{m_1}, g_{m_2})= (1, -1) $ subspace with a single dynamical matter excitation present.
For $h=0$, the dynamics are 
restricted to the two states
\begin{equation}
\label{eq:stretch_compress}
\ket{\psi(t)} = \cos(Jt) \ket{1_{m_1}, -_\ell, 0_{m_2}} + i \sin(Jt) \ket{0_{m_2}, +_\ell, 1_{m_2}},  
\end{equation}
where the ket notation only includes the dynamical matter and gauge field configuration, as the background charges stay fixed.
This is a manifestation of the $\mathbb{Z}_2$ version of Gauss's law where the movement of the charge is associated with the stretching (presence)  or compressing (absence) of an electric field line emanating from the fixed background source $ q^{\rm bg}_{m_1} = 1 $ and following the dynamical charge, which acts as a sink.

Under the full Hamiltonian in Eq.~\eqref{eq:z2_gauge_link}, i.e., $h\neq 0$, the gauge ($\bar{s}_\ell^x$) and matter ($\bar{n}_{m_1}$, $\bar{n}_{m_2}$) observables oscillate with the same frequency $\Omega_0=\sqrt{J^2 +h^2}$, thus exhibiting correlations that confirm the intertwining of gauge and matter fields
\begin{equation}\label{eq:analytic_dyn_observables}\begin{split}
    \bar{n}_{m_1}(t) &\equiv \braket{\hat{a}_{m_1}^\dagger \hat{a}_{m_1}}(t) = 1 - \frac{J^2}{\Omega_0^2} \sin^2(\Omega_0 t), \\
    \bar{n}_{m_2}(t) &\equiv \braket{\hat{a}_{m_2}^\dagger \hat{a}_{m_2}}(t) = \frac{J^2}{\Omega_0^2} \sin^2(\Omega_0 t), \\
    \bar{s}_\ell^x(t) &\equiv \braket{\hat{\sigma}^x_{\ell}}(t) = -1 + \frac{2 J^2}{\Omega_0^2} \sin^2(\Omega_0 t).
\end{split}
\end{equation}
As the electric field energy increases, stretching or compressing field lines becomes increasingly costly, which inhibits charge dynamics, the precursor to confinement in larger systems~\cite{bazavan2024synthetic}.

To add an additional spatial dimension, we need to add more links, which in turn requires a larger qubit register.
The minimal building block for $2+1$D,
is realised with one additional qubit. Both qubits encode gauge fields and independently mediate tunnelling between two matter sites forming a loop, as shown in Fig.~\ref{fig:scheme_link}(e, f). This configuration resembles a traditional Wilson loop~\cite{PhysRevD.10.2445} but involves only two matter sites instead of four.


The dynamics within the $\mathbb{Z}_2$-loop are described by
\begin{equation}\label{eq:plaquette_hamiltonian}
    \hat{H}_{\mathbb{Z}_2}^{\rm loop} = \sum_{i=1,2}\left(\left(J\hat{a}^{{\dagger}}_{m_2}\hat{\sigma}^z_{\ell_i}\hat{a}^{\phantom{\dagger}}_{m_1}+{\rm H.c.}\right)+ h\hat{\sigma}^x_{\ell_i}\right),
\end{equation}
which retains a local symmetry generated by $\hat{G}^{\phantom{x}}_{m_i} =~  \hat{\sigma}^x_{\ell_1}\hat{\sigma}^x_{\ell_2} \hat{P}_{m_i}^{\phantom{x}}$ for $i\in\{1,2\}$. 
To ensure Gauss's law, we could initialise the gauge fields in the electric-field basis, \(\ket{+_{\ell_1}, +_{\ell_1}}\), \(\ket{-_{\ell_1}, -_{\ell_1}}\), \(\ket{-_{\ell_1}, +_{\ell_1}}\), or \(\ket{+_{\ell_1}, -_{\ell_1}}\), which are eigenstates of \(\hat{G}_{m_i}\). Here, unlike in the single-link case, Gauss's law is less restrictive, allowing us to work in the magnetic-field basis instead. In this basis, the gauge fields are prepared in superpositions forming maximally entangled Bell pairs, ${\ket{\Phi^\pm_{\ell_1\ell_2}} = (\ket{+_{\ell_1},\pm_{\ell_2}}+ \ket{-_{\ell_1}, \mp_{\ell_2}})/\sqrt{2}}$ and
${\ket{\Psi^\pm_{\ell_1\ell_2}}= (\ket{\pm_{\ell_1}, +_{\ell_2}} - \ket{\mp_{\ell_1}, -_{\ell_2}}})/\sqrt{2}$~\cite{bellstates}.
These states encode the presence or absence of a $\mathbb{Z}_2$ magnetic flux through the surface enclosed by the two links. As the dynamical matter excitation tunnels via either link, it acquires a relative phase depending on the gauge field configuration. 
This results in constructive or destructive interference between the two paths leading to the observation of the Aharonov–Bohm effect, a surface ($2+1$D) phenomena, with a phase given by
\begin{equation}
\phi_{\rm AB}=\arccos(\langle\hat{\sigma}_{\ell_1}^z\hat{\sigma}_{\ell_2}^z\rangle).
\end{equation}


In the $\ket{\Phi_{\ell_1 \ell_2}^\pm}$ configuration, a matter excitation acquires the same phase whether it tunnels via $\ell_1$ or $\ell_2$ (i.e., ${\phi_{AB} = 0}$), indicating the absence of magnetic flux. initialising in ${\ket{\psi_0} = \ket{1_{m_1}, \Phi^+_{\ell_1\ell_2}, 0_{m_2}}}$ and setting $h = 0$, the state evolves as
\begin{equation}\label{eq:phi_pls_tunneling}
    \ket{\psi(t)} = \cos(2 J t)\ket{1_{m_1}, \Phi^+_{\ell_1\ell_2}, 0_{m_2}} + i \sin(2 J t)\ket{0_{m_1}, \Phi^-_{\ell_1\ell_2}, 1_{m_2}},
\end{equation}
with tunnelling at rate $2J$. 
Enforced by Gauss’s law, the \(\mathbb{Z}_2\) charge dynamics synchronise with coherent oscillations between orthogonal Bell states, producing a periodic inversion of \(\langle \hat{\sigma}_{\ell_1}^x \hat{\sigma}_{\ell_2}^x \rangle\), while the \(\mathbb{Z}_2\) magnetic flux (\(\phi_{\rm AB} = 0\)) remains zero throughout. This behaviour resembles the stretching and compression of the electric field line observed in the single-link case.

Conversely, in the $\ket{\Psi_{\ell_1 \ell_2}^\pm}$ configuration, the matter excitation acquires phases $\phi_1$ or $\phi_2 = \phi_1 + \pi$ when tunnelling via $\ell_1$ or $\ell_2$, resulting in $\phi_{AB} = \pi$ and thus magnetic flux and vison are present. This causes destructive Aharonov-Bohm interference, inhibiting tunnelling; hence, the state ${\ket{\psi_0} = \ket{1_{m_1}, \Psi_{\ell_1 \ell_2}^\pm, 0_{m_2}}}$ remains frozen for all time.

\section*{Initialising and measuring the hybrid system}
In the experiment, the matter sites are encoded in harmonic oscillators formed by normal modes of vibration of a one- or two-ion crystal. We can prepare these oscillators with $k$ quanta $\ket{k_{m_i}}$.
Meanwhile, each gauge field is represented by a qubit, encoded in the internal level structure of each ion in the crystal, with the qubit states given by two electronic levels, \( \ket{\downarrow_\ell} \) and \( \ket{\uparrow_\ell} \). Experimental details of our trapped-ion system are provided in the Supplement~\cite{supplementary}.

We first focus on realising the dynamics of a single link using a single ion, shown in Fig.~\ref{fig:scheme_link}(c,d), where the LGT degrees of freedom are encoded in its qubit $\ell$ and two motional modes, $m_1$ and $m_2$. The quantum processor operates in a hybrid mode: digital for state preparation, error mitigation, and readout, and analogue during gauge-invariant tunnelling. By applying the tunnelling interaction for a variable duration, we can simulate the time evolution of any initialised state.

To synthesise the first term in Eq.~\eqref{eq:z2_gauge_link}, the matter tunnelling mediated by the gauge field, we employ the scheme proposed in Ref.~\cite{sutherland2021universal}, previously used to generate single-mode generalised squeezing~\cite{bazavan2024squeezing, saner2024generating}. This scheme combines two qubit state-dependent forces, each addressing a different oscillator: one detuned by \(\Delta\) from oscillator \(m_1\), the other by \(\Delta\) from oscillator \(m_2\). Since the forces are conditioned on \(\hat{\sigma}_\ell^x\) and \(\hat{\sigma}_\ell^y\), their non-commutativity gives rise to an effective resonant tunnelling interaction conditioned on \(\hat{\sigma}_\ell^z \propto [\hat{\sigma}_\ell^x, \hat{\sigma}_\ell^y]\). The ability to engineer qubit state-dependent forces with arbitrary conditioning enables complete control over the state-dependence of the tunnelling term in Eq.~\eqref{eq:z2_gauge_link}. Furthermore, this scheme relies on smoothly switching the interaction on and off over a duration \(t_R\), thus setting a lower bound on the pulse width: \(t \geq t_R\). A detailed explanation of this scheme is provided in the Supplement~\cite{supplementary}. Alternative methods to create state-conditioned tunnelling exist, e.g., Ref.~\cite{gan2020hybrid,chen2023scalable}; however the resulting Hamiltonian may not be gauge-invariant.


To initialise the hybrid system in $\ket{\psi_0} = \ket{1_{m_1}, -_\ell, 0_{m_2}}$, we first cool the vibrational modes close to their ground state ($\bar{n} \approx 0$) via Doppler and resolved-sideband cooling. The ion's internal state is then optically pumped to $\ket{\downarrow_\ell}$. A resonant anti-Jaynes-Cumming interaction (blue sideband pulse) on $m_1$ prepares $\ket{1_{m_1}, \uparrow_\ell, 0_{m_2}}$, followed by a $\pi/2$ qubit rotation to produce $\ket{1_{m_1}, -_\ell,0_{m_2}}$ (see Supplement~\cite{supplementary}).


The readout sequence for the hybrid state depends on the observable. We measure the gauge field state ($\bar{s}_\ell^x$) by applying a \(\pi/2\) qubit rotation followed by fluorescence detection. To efficiently read out matter site expectation values (\(\bar{n}_{m_1}\), \(\bar{n}_{m_2}\)), we restrict the tomography to the subspace with at most one excitation per site. 
We perform fluorescence detection and post-select on the dark state, tracing out qubit-oscillator entanglement. We then read out the matter state  via the qubit by applying a blue sideband pulse on the respective oscillator and perform a final fluorescence detection (see Supplement~\cite{supplementary} or Ref.~\cite{gan2020hybrid}).



\section*{$\mathbb{Z}_2$ Link tunnelling}
\begin{figure}[ht]
    \centering
    \includegraphics[]{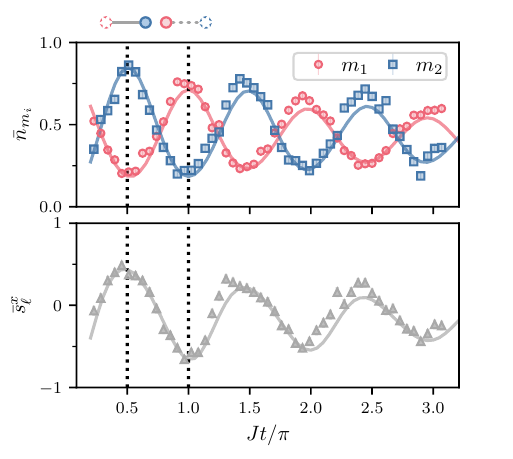}
    \caption{
    Matter excitation dynamically tunnelling between matter site $m_1$ and $m_2$ as a function of the evolution time of the Hamiltonian in Eq.~\eqref{eq:z2_gauge_link} with $h=0$. We prepare the initial state $\ket{1_{m_1}, -_\ell, 0_{m_2}}$. We measure the observables corresponding to the gauge field $\bar{s}_\ell^x$ and matter occupation $\bar{n}_{m_1}$ and $\bar{n}_{m_2}$. As the matter tunnels between sites, the gauge field inverts. 
    The insets above the dotted lines illustrate the system’s configuration at different time points, corresponding to the states $\ket{0_{m_1},+_\ell, 1_{m_1}}$ and $\ket{1_{m_1},-_\ell, 0_{m_1}}$. 
    Solid lines show simulations using independently estimated parameters, except for the tunnelling rate \( J \), which is floated to match the experimental timescale. The required 12\% correction, applied in the displayed numerics, indicates a discrepancy  relative to theoretical estimates and is left for future investigation. We extracted a tunnelling rate of $J/\pi = \SI{1.47}{\kilo\hertz}$. Due to the required pulse ramping, the minimal possible interaction duration (full-width half maximum) is given by $2t_{R} J/\pi =0.24$ instead of $0$. Error bars from quantum projection noise (68\% confidence interval) are smaller than the markers and are omitted.
    }
    \label{fig:link_tunneling}
\end{figure}

We first consider the case of vanishing electric field energy, i.e., \(h = 0\) in Eq.~\eqref{eq:z2_gauge_link}. After initialising the system in the state \(\ket{\psi_0} = \ket{1_{m_1}, -_\ell, 0_{m_2}}\), we evolve it under gauge-invariant tunnelling alone and perform the readout. Figure~\ref{fig:link_tunneling} shows the experimentally measured expectation values of the gauge-invariant observables \(\bar{n}_{m_1}(t)\), \(\bar{n}_{m_2}(t)\), and \(\bar{s}_\ell^x(t)\), alongside numerical simulations. The data demonstrate coherent charge tunnelling synchronised with the stretching and compressing of electric field lines. The agreement with numerical predictions confirms our understanding of how imperfections in the quantum processor affect the dynamics. The correlation between the three experimentally measured observables and the agreement between experiment and numerical simulation are quantified in the Supplement~\cite{supplementary} for all data presented in this and subsequent figures.
Unlike Eq.~\eqref{eq:analytic_dyn_observables}, these predictions incorporate an initial thermal occupation of \(\bar{n}_{m_1}(0) = \bar{n}_{m_2}(0) = 0.1\) due to imperfect ground-state cooling, as well as motional heating rates of \(\dot{\bar{n}}_{m_1} = \SI{300} {quanta/s}\) and \(\dot{\bar{n}}_{m_2} = \SI{5}{quanta/s}\). These independently calibrated decoherence sources explain the reduced and slowly decaying contrast compared to the idealised case, which we show in the Supplement~\cite{supplementary}. Although these errors weakly break gauge symmetry, we still observe correlated dynamics of matter and gauge fields.

Other sources of error include uncompensated AC-Stark shifts and magnetic field fluctuations, which introduce a dephasing term proportional to \(\hat{\sigma}_\ell^z\) that could also break gauge invariance. To mitigate these effects, we split the tunnelling dynamics into two parts, interleaved with a qubit-echo refocusing sequence~\cite{supplementary,hahn1950spin}.

\section*{$\mathbb{Z}_2$ link tunnelling and electric field energy}
\begin{figure*}
    \centering    \includegraphics[width=1\linewidth]{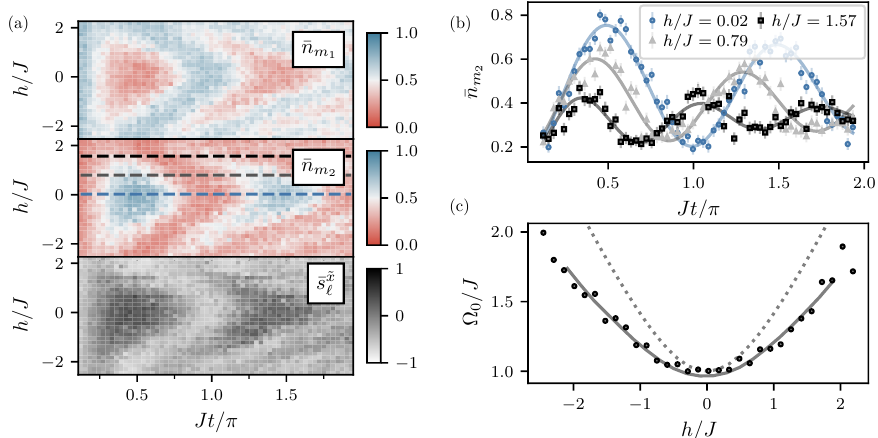}
    \caption{Real-time dynamics of the observables $\bar{n}_{m_1}(t)$, $\bar{n}_{m_2}(t)$, and $\bar{s}_\ell^{\tilde{x}}(t)$ evolving under the full $\mathbb{Z}_2^{\rm link}$ Hamiltonian given by Eq.~\eqref{eq:z2_gauge_link} but in the rotated basis (see main text). 
    (a) In the two-dimensional scans, we find clear correlations between matter ($\bar{n}_{m_1}$, $\bar{n}_{m_2}$) and gauge field observable ${\bar{s}_\ell^{\tilde{x}} \equiv \bar{s}_\ell^z}$, where $\tilde{x}$ takes the role of the $x$ observable in the transformed basis. (b) We display three slices of $\bar{n}_{m_2}$ at electric field settings $h/J = \{0.02, 0.79, 1.57\}$ as indicated by the dashed lines in (a). We fit a decaying sinusoid to each of the slices. Error bars denote the $68\%$ confidence interval due to quantum projection noise. (c) We show the extracted oscillation frequency $\Omega_0/J$ versus the field energy $h/J$. We measure a tunnelling rate $J/\pi = \Omega_0(0)/\pi =  \SI{1.510(6)}{\kilo\hertz}$. The experimental data agrees with the numerical simulation using independent parameter estimates (solid line). We do, however, find a deviation from the analytic theory $\Omega_0/J = \sqrt{1 +(h/J)^2}$ (dotted line), which indicates that the
    the effective electric field term that is introduced has a smaller magnitude than anticipated from the analytic expression.}
    \label{fig:tunneling_with_h_term}
\end{figure*}

Next, we want to simulate the dynamics in Eq.~\eqref{eq:z2_gauge_link} with electric field energy $|h| > 0$. To minimise technical overhead, we implement the tunnelling conditioned on $\hat{\sigma}_\ell^x$, which allows us to add the electric field term by introducing a qubit frequency offset $h\hat{\sigma}_\ell^z$. This yields the Hamiltonian in Eq.~\eqref{eq:z2_gauge_link}, but in a rotated basis, where $\hat{\sigma}_\ell^{\tilde{z}} = \hat{\sigma}_\ell^x,\ \hat{\sigma}_\ell^{\tilde{x}} = \hat{\sigma}_\ell^z$ (see Supplement~\cite{supplementary}). 
We investigate the dynamics of the matter excitation and the correlated gauge field, while keeping $J$ constant and varying $h$, as shown in Fig.~\ref{fig:tunneling_with_h_term}(a).

We extract $\Omega_0(h)$ by fitting a decaying sinusoid to the time evolution of $\bar{n}_{m_2}(t)$, as shown in Fig.~\ref{fig:tunneling_with_h_term}(b,c). Consistent with the prediction of Eq.~\eqref{eq:analytic_dyn_observables}, we observe that increasing $|h|$ enhances the tunnelling rate $\Omega_0$ while suppressing the oscillation amplitude. The extracted values of $\Omega_0$ agree with numerical simulations but deviate from the analytic theory, suggesting that the effective electric field term is smaller than expected.

The observed dynamics indicate a gradual transition from a tunnelling-dominated regime to one where the gauge field becomes effectively pinned, inhibiting matter tunnelling, as $|h/J|$ increases. This is a direct manifestation of the intertwining of charge and gauge-field properties:
stretching or compressing the electric field line becomes increasingly costly, and hence, charge tunnelling is suppressed.
In larger lattices, $|h|>0$ would lead to a linearly rising potential between a pair of charges, resulting in a Wannier-Stark confinement~\cite{bazavan2024synthetic} analogous to the quark-anti-quark confinement in mesons~\cite{bazavan2024synthetic}.

\section*{Aharonov-Bohm effect on a $\mathbb{Z}_2$ loop}
We introduce a second spatial dimension to our system to observe real-time dynamics in a \(\mathbb{Z}_2\) loop shown in Fig.~\ref{fig:scheme_link}(e,f) and described by the gauge-invariant Hamiltonian in Eq.~\eqref{eq:plaquette_hamiltonian}. This is achieved by adding a second ion to the crystal, which introduces an additional gauge field that independently mediates tunnelling between the two matter sites. The matter sites $(m_1, m_2)$ correspond to two of the normal motional modes of the two-ion crystal. To simulate the effect of magnetic flux piercing the loop, we initialise the gauge fields $(l_1,l_2)$ in the magnetic field basis.  
Using the digital control capabilities of the trapped-ion hybrid device, we prepare two initial entangled gauge field configurations, \(\ket{\psi_0} \in \left\{ \ket{1_{m_1},\Phi^+_{\ell_1 \ell_2},0_{m_2}}, \ket{1_{m_1},\Psi^+_{\ell_1 \ell_2},0_{m_2}} \right\}\), where \(\Phi_{\ell_1\ell_2}^+\) and \(\Psi_{\ell_1\ell_2}^+\) are Bell states generated through a sequence of single- and two-qubit entangling gates (see Supplement~\cite{supplementary}). These gauge field states correspond to effective magnetic fluxes of $0$ and \(\pi\) through the loop, respectively.
\begin{figure}[ht!]
    \centering
    \includegraphics[width=1\linewidth]{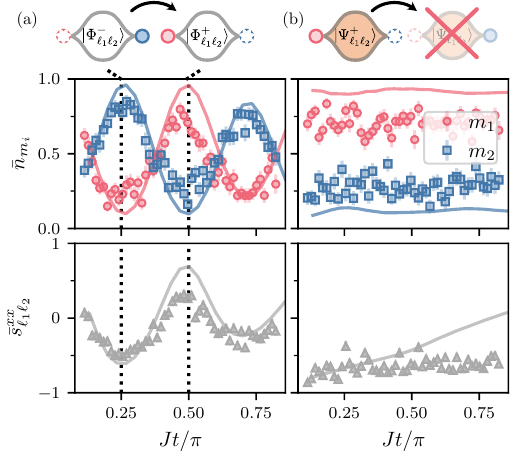}
    \caption{
tunnelling of matter particles under Aharonov-Bohm interference in the $\mathbb{Z}_2$ loop (Eq.~\eqref{eq:plaquette_hamiltonian}). By preparing initial states with entangled gauge fields, we emulate a magnetic flux and encode an Aharonov-Bohm phase $\phi_{AB}$ via the gauge fields’ dynamical degrees of freedom. We track the expectation values $\bar{n}_{m_1}$, $\bar{n}_{m_2}$, and $\bar{s}_{\ell_1\ell_2}^{xx}$.
(a) For the initial state $\ket{1_{m_1}, \Phi_{\ell_1\ell_2}^+, 0_{m_2}}$, corresponding to zero magnetic flux ($\phi_{AB} = 0$), we observe coherent matter tunnelling correlated with $\bar{s}_{\ell_1\ell_2}^{xx}$. The insets above the dotted lines show the intermediate states $\ket{0_{m_1}, \Phi_{\ell_1\ell_2}^-, 1_{m_2}}$ and $\ket{1_{m_1}, \Phi_{\ell_1\ell_2}^+, 0_{m_2}}$. Supplementary data confirm oscillations between these configurations~\cite{supplementary}. We extract a tunnelling rate of $J/\pi = \SI{0.35}{\kilo\hertz}$.
(b) For $\ket{1_{m_1}, \Psi_{\ell_1\ell_2}^+, 0_{m_2}}$, shown in the inset, a $\pi$ flux ($\phi_{AB} = \pi$, orange shade) is introduced, leading to destructive Aharonov-Bohm interference, where tunnelling is inhibited. All observables remain constant.
Solid lines are numerical simulations using independently estimated parameters including oscillator heating and qubit decoherence, except for the tunnelling rate \( J \), which is floated to match the experimental timescale. Error bars from quantum projection noise (68\% confidence) are comparable to the marker size.
    }
    \label{fig:plaqutte_ab_effect}
\end{figure}

We then apply the analogue tunnelling interaction to observe the real-time evolution of the \(\mathbb{Z}_2\) loop under the gauge-invariant Hamiltonian~\eqref{eq:plaquette_hamiltonian} with $h=0$. Following this, we measure system observables: the matter site occupations \(\bar{n}_{m_1}(t)\) and \(\bar{n}_{m_2}(t)\), as well as the gauge-field correlator \(\bar{s}_{\ell_1 \ell_2}^{xx}(t) = \braket{\hat{\sigma}^x_{\ell_1} \hat{\sigma}^x_{\ell_2}}\), using the same methods described in the $\mathbb{Z}_2$ link simulation. 

These observables, shown in Fig.~\ref{fig:plaqutte_ab_effect}, track the correlated dynamics between matter excitations and the entangled gauge fields. Additionally, we verify in the Supplement~\cite{supplementary} that \(\bar{s}_{\ell_1 \ell_2}^{zz}(t) = \braket{\hat{\sigma}^z_{\ell_1} \hat{\sigma}^z_{\ell_2}}\) remains constant, confirming that the flux and corresponding Aharonov-Bohm phase \(\phi_{\rm AB} = \arccos\bigl(\braket{\hat{\sigma}^z_{\ell_1} \hat{\sigma}^z_{\ell_2}}\bigr)\) are conserved throughout the evolution, aside from a gradual amplitude decay caused by decoherence.

When the gauge fields are initialised in the Bell state \(\ket{\Phi_{\ell_1 \ell_2}^+}\), corresponding to zero magnetic flux \((\phi_{\rm AB} = 0)\), no vison is present. In this scenario (Fig.~\ref{fig:plaqutte_ab_effect}(a)), the matter excitation tunnels freely, with dynamics intertwined with the gauge field, oscillating between \(\ket{1_{m_1}, \Phi_{\ell_1 \ell_2}^+, 0_{m_2}} \leftrightarrow \ket{0_{m_1}, \Phi_{\ell_1 \ell_2}^-, 1_{m_2}}\), as described by Eq.~\eqref{eq:phi_pls_tunneling}. In contrast, the Bell state \(\ket{\Psi_{\ell_1 \ell_2}^+}\) corresponds to a flux \(\phi_{\rm AB} = \pi\), indicating the presence of a vison. This flux induces destructive interference that suppresses tunnelling (Fig.~\ref{fig:plaqutte_ab_effect}(b)), effectively freezing the system in the initial configuration.




\section*{Future work}
Our experimental findings demonstrate for the first time the role of gauge-field entanglement in Aharonov-Bohm interference, and the weaving of flux and charge, in a LGT quantum simulator. The advances presented here highlight the potential of the hybrid qubit-oscillator encoding for simulating LGTs in synthetic dimensions. Building on this, we outline three directions for future exploration.
\begin{figure}[ht]
    \centering
    \includegraphics[width=1\linewidth]{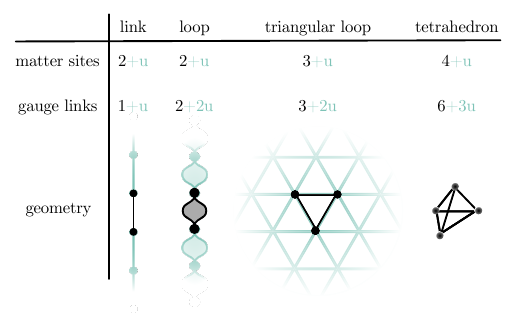}
    \caption{
    Resource requirements for elementary building blocks. The link and loop blocks, which have been realised in this work, as well as the triangular loop and tetrahedron blocks, which require only a modest increase in register size, are shown in black. The lattices formed by concatenating these blocks are illustrated in green. Resource scaling of a lattice formed by $u+1$ blocks is also indicated.
    }
    \label{fig:extending_geometries}
\end{figure}
Firstly, introducing an electric field energy ($|h|>0$) on the loop leads to a three-way competition between the Aharonov-Bohm effect, tunnelling, and the field energy itself. This interplay offers a tunable setting for exploring interference-driven localization and its breakdown in a gauge-invariant context.
We provide numerical simulations of the ideal dynamics under this additional interaction (see Supplement \cite{supplementary}). 

Secondly, the bosonic nature of the matter sites, encoded in the harmonic oscillators, enables richer matter dynamics that are not constrained by the Pauli exclusion principle.
Beyond the single-phonon matter excitations discussed so far, this approach allows us to explore the dynamics of superpositions of Fock states with well-defined \(\mathbb{Z}_2\) charges. We focus in particular on squeezed matter states, which we investigate both numerically and experimentally, as detailed in the Supplement~\cite{supplementary}. This direction is promising, as the Hilbert space dimension grows with the squeezing magnitude, potentially enabling classically intractable real-time dynamics without requiring large lattice sizes. Additionally, it allows for novel characterizations
using phase-space methods, including direct measurement of the characteristic function~\cite{fluhmann2020direct,PhysRevLett.133.050602,bazavan2024squeezing}.

Finally, synthetic dimensions open new possibilities for exploring LGTs
in complex geometries. Engineering tunnelling between specific motional modes (representing matter sites) by addressing individual ions (representing links) enables various configurations, such as chains of connected loops~\cite{domanti2024dynamical}, triangular loops~\cite{homeier2021z2}, and even tetrahedral lattices, the simplest 
instance of a $3+1$D LGT, all with modest experimental resources (Fig.~\ref{fig:extending_geometries}).

In conclusion, we have experimentally demonstrated a \(\mathbb{Z}_2\) lattice gauge theory on a single link, encoding dynamical matter and gauge fields on equal footing within a hybrid qubit-oscillator trapped-ion system. We employ digital techniques to prepare initial states and read out observables, while analogue evolution simulates real-time dynamics. Utilising synthetic dimensions, we extended this to an elementary \(2+1\)D geometry, achieving the first quantum simulation of the real-time dynamics of a \(\mathbb{Z}_2\) loop. By preparing entangled gauge-field states corresponding to effective magnetic fluxes of \(0\) and \(\pi\), we observed the Aharonov–Bohm effect. Matter tunnelling was inhibited for \(\phi_{\rm AB} = \pi\), confirming the presence of a vison, while for \(\phi_{\rm AB} = 0\), tunnelling remained uninhibited.

We have outlined several directions for extending this work: probing the competition between magnetic flux and electric field energy, exploring bosonic matter dynamics, and engineering more complex lattice geometries, enabled by the hybrid processor’s digital-analogue control and qubit-oscillator architecture. More broadly, these capabilities provide a rich foundation for further theoretical and experimental progress, paving the way for the exploration of novel states of matter and phase transitions in larger Hilbert spaces with flexible connectivity, beyond the reach of classical methods.


\section*{Acknowledgments}
We would like to thank Peter Drmota, for insightful discussions on converting between different Bell states, David Nadlinger for his valuable input on micromotion-sideband-based single addressing, and Kai Shinborough and Iver \O{}vergaard for feedback on the manuscript. 
This work was supported by the US Army Research Office (W911NF-20-1-0038) and the UK EPSRC Hub in Quantum Computing and Simulation (EP/T001062/1). GA acknowledges support from Wolfson College, Oxford. CJB acknowledges support from a UKRI FL Fellowship. RS acknowledges funding from an EPSRC Fellowship EP/W028026/1 and Balliol College, Oxford. AB
acknowledges support from PID2021-127726NB- I00 (MCIU/AEI/FEDER, UE),
from the Grant IFT Centro de Excelencia Severo Ochoa CEX2020-001007-S,
funded by MCIN/AEI/10.13039/501100011033, from the CSIC Research
Platform on Quantum Technologies PTI-001, and from the European Union’s
Horizon Europe research and innovation programme under grant agreement
No 101114305 ("MILLENION-SGA1” EU Project).

\bibliography{bibliography}

\begin{thebibliography}{58}
\bibitem [{bel()}]{bellstates}%
  \BibitemOpen
  \href@noop {} {}\bibinfo {note} {N.b., these are the regular Bell states that we typically express in the computational basis $\ket{\Phi_{\ell_1\ell_2}^\pm} = ( \ket{\uparrow_{\ell_1},\uparrow_{\ell_2}} \pm \ket{\downarrow_{\ell_1},\downarrow_{\ell_2}} )/\sqrt{2}$ and $\ket{\Psi_{\ell_1\ell_2}^\pm} = ( \ket{\uparrow_{\ell_1},\downarrow_{\ell_2}} \pm \ket{\downarrow_{\ell_1},\uparrow_{\ell_2}} ) /\sqrt{2}$.}\BibitemShut {Stop}%
\end{thebibliography}

\clearpage
\newcounter{mainfigs}
\setcounter{mainfigs}{\value{figure}}  
\renewcommand{\thefigure}{B.\number\numexpr\value{figure}-\value{mainfigs}\relax}

\newcounter{maineqs}
\setcounter{maineqs}{\value{equation}}  

\renewcommand{\theequation}{A.\number\numexpr\value{equation}-\value{maineqs}\relax}

\newcounter{mainsecs}
\setcounter{mainsecs}{\value{section}}

\newcounter{offsetsec}
\renewcommand{\thesection}{%
  \setcounter{offsetsec}{\numexpr\value{section}-\value{mainsecs}\relax}%
  \Roman{offsetsec}%
}
\onecolumngrid
\begin{center}
\vspace{5 mm}
\textbf{\large Supplemental Material for:
Real-Time Observation of Aharonov-Bohm Interference in a $\mathbb{Z}_2$ Lattice Gauge Theory on a Hybrid Qubit-Oscillator Quantum Computer}
\end{center}
\twocolumngrid
\section{Encoding of the LGT in a hybrid qubit-oscillator system}
As proposed in Ref.~\cite{bazavan2024synthetic} and discussed in the main text, we encode the gauge fields and matter sites in distinct components of a hybrid qubit-oscillator system realised with trapped ions. The trapped ion system used in this work is described in detail in Ref.~\cite{thirumalai2019high, bazavan2024easy,saner2025quantum}. The gauge field is represented by a qubit formed from two internal electronic states of a single $^{88}{\rm Sr}^+$ ion
\begin{equation}
    \ket{\downarrow_\ell} \equiv \ket{5S_{1/2}, m_j = -1/2}, \quad
\ket{\uparrow_\ell} \equiv \ket{4D_{5/2}, m_j = -3/2}.
\end{equation}
The matter sites are encoded in the normal modes of the ion crystal. To represent a link, we use a single ion, requiring only two of its three available motional modes: the axial mode with $f_{\text{ax-ip}} = \SI{1.2}{\MHz}$, heating rate $\dot{\bar{n}}_{\text{ax-ip}} =\SI{300}{quanta/s}$ and coherence time $t_{c,\text{ax-ip}} = \SI{1.7}{\milli\second}$; the lower radial mode with $f_{\text{lr-ip}} = \SI{1.5}{\MHz}$, heating rate $\dot{\bar{n}}_{\text{lr-ip}}= \SI{5}{quanta/s}$ and coherence time $t_{c,\text{lr-ip}} = \SI{2.3}{\milli\second}$; and the upper radial mode $f_{\text{ur-ip}} = \SI{1.6}{\MHz}$, heating rate $\dot{\bar{n}}_{\text{ur-ip}}= \SI{17}{quanta/s}$ and coherence time $t_{c,\text{ur-ip}} = \SI{3.3}{\milli\second}$.
The specific modes used vary across the experiments presented in this article and are listed in Table~\ref{tab:exp_params}.
For the loop configurations, we trap a two-ion crystal, which introduces an additional qubit to encode the second gauge field. The collective motion of the two ions hybridises to form six normal modes, from which we select two of the in-phase (ip) modes to represent the matter sites.

We coherently manipulate the qubit–oscillator system using a \SI{674}{\nano\meter} laser with a waist size of \SI{20}{\micro\meter} at the ion. This laser drives both the qubit transition and the qubit–oscillator interactions.

\begin{table*}[h]
\centering
    \caption{Experimental parameters for the real-time dynamics. All interactions embedded in a spin echo sequence have the duration $t$ given as $2\ {\rm x}\ t_{\rm arm}$ where $t_{\rm arm}$ is the full-width half maximum duration for which the interaction is applied in each arm. We show the used power $P$ at the ion per SDF, the strength of the SDF $\Omega_{\alpha,\alpha'}$, detuning $\Delta$ of the SDFs used to generate the interaction, the spin-conditioning $\hat{\sigma}_\beta$ and the motional modes (ax-ip: axial in-phase, lr-ip: lower radial in-phase, ur-ip: upper radial in-phase mode of the ion crystal) used to encode the matter sites.}
       \label{tab:exp_params}
        \begin{tabular}{lccccccccc} 
            Fig. &{P (mW)} & $\Omega_{\alpha},\, \Omega_{\alpha'}$ ($2\pi\, \unit{\kilo\hertz}$)& $\Delta/2\pi$ (kHz) & $t$ (\unit{\micro \second}) & $\hat{\sigma}_{\ell}^\beta$ & modes\\ \midrule
            \ref{fig:link_tunneling} & 2 & 1.201(4), 1.216(4) & -25 & 2 x (80 - 1080) & $\hat{\sigma}_{\ell}^z$ & ax-ip/lr-ip\\
            \ref{fig:tunneling_with_h_term} & 3.5 &  0.1900(7), 1.608(6), & -20 & 80 - 1280 & $\hat{\sigma}_{\ell}^y$ & ax-ip/lr-ip\\
            \ref{fig:plaqutte_ab_effect}& 2 &  1.201(4), 1.216(4) & 20/-20 & 2 x 2 x (80 - 580) & $\hat{S}_{\ell_1\ell_2}^z$ & lr-ip/ur-ip\\
            \ref{fig:supp_plus_sqz00_sx_exp}/\ref{fig:supp_plus_sqz10_sx_exp} squeezing & 2 & 1.201(4), 1.216(4) & 25 &530 & $\hat{\sigma}_{\ell}^z$ & lr-ip\\
            \ref{fig:supp_plus_sqz00_sx_exp}/\ref{fig:supp_plus_sqz10_sx_exp} tunnelling & 2 & 1.201(4), 1.216(4) & -25 & 2 x (80 - 1080) & $\hat{\sigma}_{\ell}^z$ & lr-ip/ur-ip\\
            \ref{fig:supp_measure_sqz_mag_vs_duration} squeezing & 2 & 1.201(4), 1.216(4) & 50 & 100 & $\hat{\sigma}_{\ell}^z$ & ax-ip\\
            \ref{fig:supp_measure_sqz_mag_vs_duration} tunnelling & 2 & 1.201(4), 1.216(4) & -25 & 0 - 830 & $\hat{\sigma}_{\ell}^z$ & ax-ip/lr-ip\\
        \end{tabular}
\end{table*}

\section{Microscopic interaction creating the gauge-matter tunneling}
\subsection{Link}
In this section, it is convenient to describe the microscopic interaction in terms of qubits (used to encode the gauge fields) and motional harmonic oscillators (used to encode the matter sites). To synthesise a tunnelling interaction conditioned on the qubit (gauge field) state (see Eq.~\eqref{eq:z2_gauge_link}), we follow the theoretical proposal presented in Ref.~\cite{sutherland2021universal}, which realises a universal set of qubit-conditioned continuous-variable quantum computing (CVQC) interactions. We have previously used this technique to demonstrate qubit-conditioned generalised single-mode squeezing~\cite{bazavan2024squeezing, saner2024generating}. Here, we extend the same approach to engineer a qubit-conditioned oscillator-oscillator interaction, enabling tunneling conditioned on the qubit state.

We create two qubit-dependent forces (SDFs) via the M\o{}lmer–S\o{}rensen (MS) scheme, using two bichromatic fields. Each SDF is detuned by \(\Delta\) from one of the motional modes. The MS scheme allows us to control the qubit-conditioning ($\hat{\sigma}_\ell^i$) to be $i\in \{x,y,z\}$~\cite{roos2008ion, bazavan2022synthesizing}, enabling us to set the SDFs such that they have non-commuting qubit operators. The resulting microscopic Hamiltonian, expressed in the interaction picture and after applying the rotating wave approximation, is given by:
\begin{equation}
\begin{split}
    \hat{H}_{\rm micro} = &\frac{\hbar \Omega_1}{2} (\hat{a}_{m_1}e^{-i\Delta t}+ \hat{a}_{m_1}^\dagger e^{i\Delta t}) \hat{\sigma}_{\ell}^i \mathbb{I}_{m_2}\\
    &+\frac{\hbar \Omega_2}{2} \mathbb{I}_{m_1} \hat{\sigma}_{\ell}^j (\hat{a}_{m_2} e^{-i\Delta t} +\hat{a}_{m_2}^\dagger e^{i\Delta t}),
\end{split}
\end{equation}
where $i,j \in \{x,y,z\}$.

The effective Hamiltonian governing the system’s evolution is given by the leading-order term in the Magnus expansion:
\begin{align}
    \hat{H}_{\mathbb{Z}_2}^{\rm link} &= \frac{\hbar\Omega_{\rm eff}}{2} \overbrace{[\hat{\sigma}_\ell^i,\hat{\sigma}_\ell^j]}^{2\varepsilon_{ijk}\hat{\sigma}_{\ell}^k} (\hat{a}_{m_1} \hat{a}_{m_2}^\dagger + \hat{a}_{m_1}^\dagger \hat{a}_{m_2}),\\
    \Omega_{\rm eff} &= \frac{\Omega_1 \Omega_2}{2 \Delta},
\end{align}
where $k \in \{x,y,z\}$
Besides this leading term, several other effects should be considered. First, each qubit-dependent force (SDF) gives rise to a geometric phase and circular excursions in phase space. By choosing a sufficiently long ramp duration, \( t_{\mathrm{ramp}} \gg 2\pi/\Delta \), the phase space excursions can be significantly suppressed and made negligible. Second, the geometric phase for a single qubit is global and thus physically irrelevant. However, as we discuss in the next section, this is no longer true for the loop case, which is encoded using two qubits. 

Setting $\hbar=1$, $i=x,\ j=y$ results in $k=z$, and associating $\Omega_{\rm eff}= J$ we obtain the first term in Eq.~\eqref{eq:z2_gauge_link}. This setting was used to simulate the results presented in Fig.~\ref{fig:link_tunneling} and Fig.~\ref{fig:plaqutte_ab_effect}. By switching the indices $i=y$ and $j=x$ we can conjugate the sign of the spin operator $-\hat{\sigma}_\ell^z$. This property is used to embed the interaction in a spin-echo sequence to refocus the noise without cancelling the tunnelling interaction.

To conduct the experiment in Fig.~\ref{fig:tunneling_with_h_term} we set $i=z,\ j=y$ which results in tunnelling conditioned on $k=x$. We can then introduce the electric field energy by detuning all applied optical fields relative to the qubit by $h$ giving rise to
\begin{equation}
    H_{\mathbb{Z}_2}^{\rm link'} =  \left(J\hat{a}_{m_1}^\dagger\hat{\sigma}_\ell^x \hat{a}_{m_2} + \mathrm{H.c.}\right) + h \hat{\sigma}_\ell^z,
\end{equation}
which corresponds to the full Hamiltonian in Eq.~\eqref{eq:z2_gauge_link} in the Hadamard basis $\hat{\sigma}_\ell^{\tilde{z}} = \hat{\sigma}_\ell^{x}$, $\hat{\sigma}_\ell^{\tilde{x}}=\hat{\sigma}_\ell^{z}$.
The symmetry operators in this basis read $\hat{\tilde{G}}_{m_i}= \hat{\sigma}_\ell^{\tilde{x}} \hat{P}_{m_i} = \hat{\sigma}_\ell^{z} \hat{P}_{m_i}$.

\subsection{Loop - Generalising the SDF method to two ions}
Adding a second ion to the crystal introduces a challenge: both qubits couple simultaneously to the driving field (i.e., \(\hat{\sigma}_{\ell}^k \rightarrow \hat{S}_{\ell_1\ell_2}^k = \hat{\sigma}_{\ell_1}^k \otimes \mathbb{I}_2 + \mathbb{I}_2 \otimes \hat{\sigma}_{\ell_2}^k\)). When we apply our nonlinear oscillator interactions to two ions, qubit-dependent forces (SDFs) detuned by \(\Delta\) from their respective motional modes generate a geometric phase \(\varphi_{\text{geom}} \propto (\eta \Omega_\alpha)^2 / \Delta\). For a single ion, this phase is global and can be neglected. However, with two ions, it leads to unwanted qubit-qubit entanglement.
To cancel this effect, we split the interaction pulse into two segments and reverse the detuning in the second pulse, \(\Delta \rightarrow -\Delta\). This reversal changes the direction of the phase-space trajectory, cancelling the accumulated geometric phase. Simultaneously, we adjust the interaction phase to ensure that the tunnelling contributions add constructively. To suppress any residual asymmetry caused by the detuning reversal, we embed this two-pulse sequence within each arm of a qubit-echo Ramsey sequence, such that each arm consists of two pulses: one with detuning \(\Delta\) and one with \(-\Delta\).

\section{State preparation and measuring the observables}
\subsection{Link}

\begin{figure*}
    \centering
    \includegraphics[]{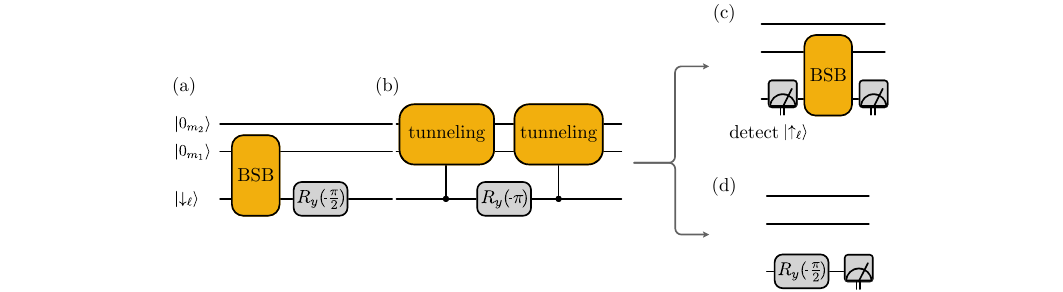}
    \caption{Circuit diagram for $\mathbb{Z}_2$ link. (a) Preparing the initial state. A blue-sideband (BSB) $\pi$ pulse prepares a single excitation in the mode it is applied to. After the following $\pi/2$-pulse on the qubit we prepare the state $\ket{1_{m_1}, -_\ell,0_{m_2}}$. (b) The $\mathbb{Z}_2$ interaction (conditioned on $\hat{\sigma}_\ell^z$ represented by a dot) is applied in two blocks split by a $\pi$-pulse. This additional $\pi$-pulse protects us against external noise $\propto \hat{\sigma}_\ell^z$ (spin-echo sequence). We adjust the phase of the second $\mathbb{Z}_2$ interaction such that the conditioning is $\propto -\hat{\sigma}_\ell^z$ such that the echo does not affect the main interaction. Finally, we apply the readout sequence. Depending on whether we read out motion or spin we apply sequence (c) or (d). (c) Readout sequence for the motion $\bar{n}_{m_1}$ and $\bar{n}_{m_2}$, respectively. We first perform a mid-circuit detection on the spin. This yields $\ket{\uparrow_\ell}$ in $50\%$ of the cases which we post-select. We can then apply a final ${\rm BSB}_{m_i}$ $\pi$-pulse to probe for an excitation in this mode. (d) Readout sequence for the spin $\bar{s}_\ell^x = \braket{\hat{\sigma}_\ell^x}$. A single $\pi/2$ rotation is followed by the projective measurement.}
    \label{fig:supp_quantum_circuit}
\end{figure*}

All motional modes used for the encoding are prepared in thermal states with mean occupation \(\bar{n} \approx 0.1\), close to but not exactly in the ground state, using Doppler and sideband cooling. To keep the notation simple, we represent the states as pure states, i.e., \(\ket{0_{m_1}, \downarrow_\ell, 0_{m_2}}\), while fully incorporating the finite temperature effects in the numerical simulations. The quantum circuit diagram of the whole experiment is depicted in Fig.~\ref{fig:supp_quantum_circuit}. We apply a blue sideband $\pi$-pulse to add a single phonon to one of the matter sites, preparing the state $\ket{1_{m_1}, \uparrow_\ell, 0_{m_2}}$. Then, we prepare the gauge field by applying a carrier $\pi/2$-pulse, reaching the desired initial state $\ket{1_{m_1}, -_\ell, 0_{m_1}}$.
For the readout, the sequence depends on the to-be-measured observable. For the gauge field observable $\bar{s}_\ell^x$, we apply another $\pi/2$ pulse and measure the spin in the computational basis. For the matter sites, we first perform a mid-circuit measurement of the spin, which collapses with equal probability to $\ket{\uparrow_\ell}$ and $\ket{\downarrow_\ell}$ and disentangles the spin from the oscillator. The qubit measurement is realised by observing the presence ($\ket{\downarrow_\ell}$) or absence ($\ket{\uparrow_\ell}$) of scattered light. Only for the absence of scatter light, i.e., $\ket{\uparrow_\ell}$ this process is non-destructive to the motional state of the ion, which encodes the matter sites. Hence, we only select the $\ket{\uparrow_\ell}$ outcomes. We then apply another blue sideband resonant with mode 1 (mode 2) to measure $\bar{n}_{m_1}$ ($\bar{n}_{m_2}$). This readout process assumes that population never leaks to Fock states $\ket{n\geq 2}$. This is not necessarily a good assumption as the axial motional mode has a large heating rate with $\dot{\bar{n}}_{m_1}=300\ {\rm quanta/s}$, which is significant given the $\sim$ms timescale of the quantum simulation. However, this only reduces the contrast of the observed tunnelling and is not detrimental to the qualitative behaviour of the simulation. Further, the heating rate can be improved with technical upgrades to the system, meaning this is not a limitation of the method.

\subsection{Loop - preparing initial states}

For the $\mathbb{Z}_2$-link, we were able to prepare Fock states via blue sideband transitions. However, for two ions, this approach breaks down because our laser addresses both qubits globally, with equal coupling strength. As a result, the blue sideband transition entangles both qubits with the oscillator, preventing the preparation of well-defined Fock states.

To overcome this, we first temporarily shelve ion 1 out of the qubit manifold by transferring the state $\ket{\downarrow_{\ell_1}}$ to an auxiliary level, $\ket{\uparrow_{\rm aux}} = \ket{S_{1/2}, m_j = 1/2}$. This places ion 1 in a spectrally isolated state that no longer couples to the laser. As a result, only ion 2 interacts with the optical field, allowing us to apply the techniques described for a single ion to prepare the desired motional Fock state. Finally, we reverse the shelving, obtaining the target state $\ket{1_{m_1}, \uparrow_{\ell_1}, \uparrow_{\ell_2}, 0_{m_2}}$.

Shelving is enabled by utilising axial micromotion sidebands that arise in our trap when ions are displaced from the RF null, which naturally occurs in a two-ion crystal. By slightly shifting the axial trap position, we create an asymmetry such that ion 1 experiences a micromotion sideband coupling of the qubit transition twice as strong as ion 2. Using this asymmetry, we first prepare the qubit configuration $\ket{\downarrow_{\ell_1}, \uparrow_{\ell_2}}$~\cite{navon2013addressing}. We then apply an RF $\pi$-pulse that transfers ion 1 from $\ket{\downarrow_{\ell_1}}$ to the auxiliary shelving state $\ket{\uparrow_{\rm aux}}$, effectively isolating it from further optical interaction. Importantly, the induced micromotion imbalance does not significantly reduce the coherence time in our system.
\begin{figure}
    \centering
    \includegraphics[width=1\linewidth]{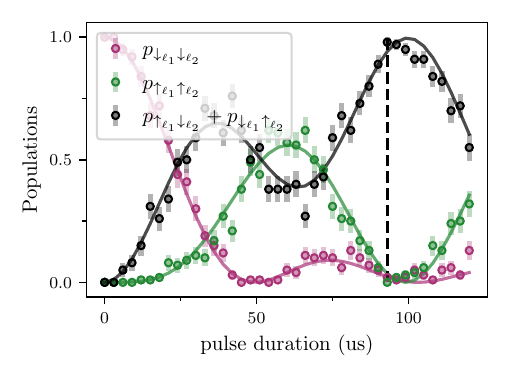}
    \caption{Evolution of the two ion populations as the micromotion sideband is driven for a variable duration. An asymmetric state $\ket{\downarrow_{\ell_1},\uparrow_{\ell_2}}$ with probability $p=0.98$ is reached after a duration of $\SI{93}{\micro\second}$, indicated by vertical line. Solid lines are a fit to the data floating the respective Rabi frequency.
    Note, that despite the readout not distinguishing between $\ket{\downarrow_{\ell_1},\uparrow_{\ell_2}}$ and $\ket{\uparrow_{\ell_1},\downarrow_{\ell_2}}$ we can be certain that we prepare only one of them (convention here: $\ket{\downarrow_{\ell_1},\uparrow_{\ell_2}}$) but not a superposition between them as the micromotion sideband drive is not producing entanglement. Error bars from quantum projection noise (68\% confidence) are indicated.}\label{fig:supp_preparing_asymmetric_state}
\end{figure}

\begin{figure}
    \centering
    \includegraphics[width=1\linewidth]{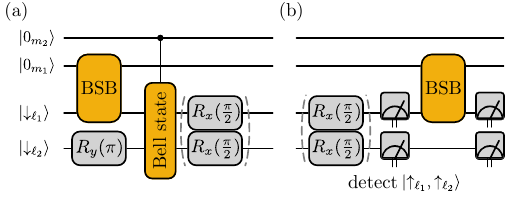}
    \caption{Simplified quantum circuit to prepare and read out the loop states. (a) The preparation sequence from Fig.~\ref{fig:supp_quantum_circuit}(a) is followed by the Bell-state creation. We use the oscillator mode in the groundstate $\ket{0_{m_2}}$ to mediate the qubit-qubit entanglement. In parentheses is the optional $R_x(\pi/2)$ rotation that is needed to go from $\Phi_{\ell_1\ell_2}^+$ to $\Psi_{\ell_1\ell_2}^+$.
    In the real sequence, we add a midcircuit measurement just before producing the Bell-state to improve the purity of the starting state. (b) For the readout of the oscillator, we convert $\Psi_{\ell_1\ell_2}^+$ back to $\Phi_{\ell_1\ell_2}^+$ ($R_x(\pi/2)$ in parentheses) and then perform a midcircuit measurement to collapse the entangled state to $\ket{\uparrow_{\ell_1},\uparrow_{\ell_2}}$. The qubit measurement is analogous to the single ion case shown in Fig.~\ref{fig:supp_quantum_circuit}(d).}
    \label{fig:supp_quantum_circuit_2ions}
\end{figure}

The quantum circuit for the state preparation sequence is shown in Fig.~\ref{fig:supp_quantum_circuit_2ions}(a). We begin by preparing the state $\ket{1_{m_1}, \uparrow_{\ell_1}, \uparrow_{\ell_2},0_{m_2}}$, and then apply a single qubit-dependent force (SDF) to entangle the gauge fields. The SDF is set near-resonant with the motional mode encoding the second matter site, which still is near the ground state. This sequence creates the Bell state $\ket{\Phi_{\ell_1\ell_2}^+}$ in $\SI{140}{\micro\second}$ with fidelity $\mathcal{F} = 0.943(9)$. Similarly, the state $\ket{\Phi_{\ell_1\ell_2}^-}$ can be prepared by starting from $\ket{1_{m_1}, \downarrow_{\ell_1}, \downarrow_{\ell_2},0_{m_2}}$.

Depending on the technique used to generate the Bell state $\ket{\Phi_{\ell_1\ell_2}^{+, \tilde\phi}} = \frac{1}{\sqrt{2}} \left( \ket{\uparrow_{\ell_1}, \uparrow_{\ell_2}} + e^{i \tilde{\phi}} \ket{\downarrow_{\ell_1}, \downarrow_{\ell_2}} \right)$, a relative phase $\tilde{\phi}$ may arise. 
It is important to track and account for \(\tilde{\phi}\) in order to correctly apply subsequent rotations in the \(\hat{\sigma}_{x}\)-\(\hat{\sigma}_y\) plane, or measure correlators, as discussed further in Sec.~\ref{sec:supp_charact_flux_and_bell}.
For example, to obtain $\ket{1_{m_1}, \Psi_{\ell_1\ell_2}^+,0_{m_2}}$, we append a global $R_x(\pi/2)$ rotation with adjusted phase offset, at the end of the preparation sequence used for $\ket{1_{m_1}, \Phi_{\ell_1\ell_2}^+,0_{m_2}}$.

\subsection{Loop - Measuring observables}
Measuring the gauge field observables generalises naturally from the single-link case. In contrast, measuring the matter sites requires additional care. Here, we again employ a mid-circuit detection scheme. This method can be directly applied to the states $\ket{\Phi_{\ell_1\ell_2}^\pm}$. In half of the measurement outcomes, no fluorescence is observed, indicating that the gauge fields have collapsed to state $\ket{\uparrow_{\ell_1}, \uparrow_{\ell_2}}$, while the oscillator state remains intact.

However, for the state $\ket{\Psi_{\ell_1\ell_2}^+}$, a direct mid-circuit measurement would always result in fluorescence from one of the qubits, thereby destroying the motional state. To avoid this, we first apply a $\pi/2$ rotation that maps the state back to $\ket{\Phi_{\ell_1\ell_2}^+}$, allowing the mid-circuit measurement to proceed without affecting the oscillator. The full protocol is illustrated in Fig.~\ref{fig:supp_quantum_circuit_2ions}(b).

\section{Quantifying correlation between experimental observables}
To quantify the correlations between the gauge field observables ($\bar{s}_\ell^x$ or $\bar{s}_{\ell_1\ell_2}^{xx}$) and the matter excitations ($\bar{n}_{m_1}$ and $\bar{n}_{m_2}$) visible in the figures, we compute the Pearson correlation coefficient, which assumes a linear relationship between the observables. We find strong correlations between observables in all three figures of the main text. Detailed results are shown in Tab.~\ref{tab:corr_coeff}.
\begin{table}[h]
\centering
    \caption{Correlation coefficients between experimentally measured observables. We show the Pearson correlation coefficients $r$ between gauge field observables ($\bar{s}_\ell^x/ \bar{s}_{\ell_1\ell_2}^{xx}$) and average matter excitation $\bar{n}_{m_i}$.}
       \label{tab:corr_coeff}
        \begin{tabular}{cccc} 
            Fig. & $r_{\ell, m_1}$ & $r_{\ell, m_2}$ & $r_{m_1, m_2}$ \\ \midrule
            \ref{fig:link_tunneling} &  $-0.94$ & $0.92$ & $-0.96$\\
            \ref{fig:tunneling_with_h_term} & $-0.78$ & $0.78$ & $ -0.84$\\
            \ref{fig:plaqutte_ab_effect}(a) & $0.88$ & $-0.88$ & $-0.87$\\
            \ref{fig:plaqutte_ab_effect}(b) & $0.17$ & $0.02$ &  $-0.19$\\
            \end{tabular}
\end{table}
\section{Experimental versus numerical data}
\subsection{Quantifying agreement}
We quantify the match between the experimentally measured and numerically simulated observables using the root mean squared error (rmse):
\begin{equation}
    \text{rmse} = \sqrt{ \frac{1}{n} \sum_{i=1}^{n} \left( y_i^{\rm exp} - y_i^{\rm num} \right)^2 }.
\end{equation}
For the observables \( n_{m_1} \) and \( n_{m_2} \), we compute the RMSE directly, as their values naturally lie within the range \([0, 1]\). For the spin observable $\bar{s}_\ell^x$/$\bar{s}_{\ell_1\ell_2}^{xx}$, we rescale it as \( (\bar{s}_\ell^x/\bar{s}_{\ell_1\ell_2}^{xx} + 1)/2 \) to map it to the same \([0, 1]\) range prior to computing the RMSE. The values for all main figure data are summarised in Tab.~\ref{tab:rmse}. We find good agreement throughout all the Figures.

\begin{table}[h]
\centering
    \caption{Root mean squared error (rmse) between experimentally measured and numerically simulated values for observables ($\bar{s}_\ell^x/ \bar{s}_{\ell_1\ell_2}^{xx}$, $\bar{n}_{m_1}$, and $\bar{n}_{m_2}$).}
       \label{tab:rmse}
        \begin{tabular}{cccc} 
            Fig. & ${\rm rmse}_{\ell}$ & ${\rm rmse}_{m_1}$ & ${\rm rmse}_{m_2}$ \\ \midrule
            \ref{fig:link_tunneling} & $0.054$ &  $0.044$ & $0.053$\\
            \ref{fig:tunneling_with_h_term} & 0.105 & 0.155 & 0.108\\
            \ref{fig:plaqutte_ab_effect}(a) & 0.106 & 0.139 & 0.107\\
            \end{tabular}
\end{table}
\subsection{Predicting expected dynamics without symmetry breaking noise}
\begin{figure}
    \centering
    \includegraphics[]{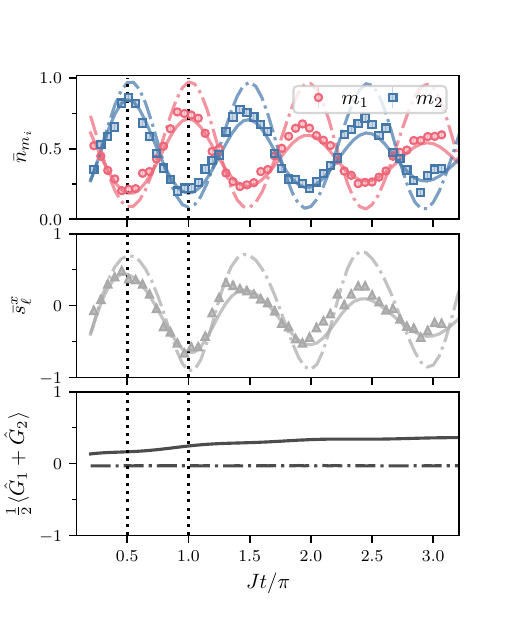}
    \caption{Additional numerical simulations accompanying Fig.~\ref{fig:link_tunneling}. Panel three shows the numerically computed expectation value of the gauge generator $\braket{\hat{G}_{m_1} + \hat{G}_{m_2}}/2$. In all panels, the solid lines correspond to numerical models that include all relevant experimental error sources and are also shown in the main figure. The dash-dotted lines represent numerical results where the oscillator mode heating, the dominant source of symmetry breaking, has been artificially set to zero. This comparison highlights the impact of oscillator heating on the simulated system dynamics and its impact on gauge invariance.}
    \label{fig:supp_fig2_with_no_heating}
\end{figure}

We know that the leading symmetry breaking error source in our system are heating effects of the oscillators. Justified by the good agreement between the numerical and experimental data, we also numerically investigate the dynamics with a heating rate reduced to zero on both modes. Additionally, we also model the expectation values of the summed symmetry operators ($\braket{\hat{G}_{m_1} + \hat{G}_{m_2}}/2$). We show this ideal case data along side the data presented in the main text in Fig.~\ref{fig:supp_fig2_with_no_heating}.

\section{In-depth investigation of real-time dynamics on the $\mathbb{Z}_2$-loop}
\subsection{Analytic derivation}
The gauge field part of the eigenstate of the gauge-symmetry operators $\hat{G}_{m_1}$ and $\hat{G}_{m_2}$ are given by the Bell states 
\begin{align}
\begin{split}
\ket{\Phi_{\ell_1\ell_2}^\pm} &= (\ket{\uparrow_{\ell_1},\uparrow_{\ell_2}}\pm \ket{\downarrow_{\ell_1} , \downarrow_{\ell_2}})/\sqrt{2} \\
&= (\ket{+_{\ell_1},\pm_{\ell_2}}+ \ket{-_{\ell_1}, \mp_{\ell_2}})/\sqrt{2},
\end{split}
\\
\begin{split}
\ket{\Psi_{\ell_1\ell_2}^\pm}  &= (\ket{\uparrow_{\ell_1}\downarrow_{\ell_2}} \pm \ket{\downarrow_{\ell_1},\uparrow_{\ell_2}})/\sqrt{2}\\
&= (\ket{\pm_{\ell_1}, +_{\ell_2}} - \ket{\mp_{\ell_1}, -_{\ell_2}})/\sqrt{2}.
\end{split}
\end{align}
For a single bosonic excitation in matter site $1$ we can then find the following operator relations

\begin{align}
    \begin{split}
    \hat{G}_{m_1} \ket{1_{m_1},\Phi_{\ell_1\ell_2}^\pm,0_{m_2}} &= (\pm 1) (-1) \ket{1_{m_1},\Phi_{\ell_1\ell_2}^\pm,0_{m_2}} \\
    &= \mp \ket{1_{m_1},\Phi_{\ell_1\ell_2}^\pm,0_{m_2}},\\
    \hat{G}_{m_2} \ket{1_{m_1},\Phi_{\ell_1\ell_2}^\pm,0_{m_2}} &= (\pm 1) (1) \ket{1_{m_1},\Phi_{\ell_1\ell_2}^\pm,0_{m_2}} \\
    &= \pm \ket{1_{m_1},\Phi_{\ell_1\ell_2}^\pm,0_{m_2}},\\
        \hat{G}_{m_1} \ket{1_{m_1},\Psi_{\ell_1\ell_2}^\pm,0_{m_2}} &= (\pm 1) (-1) \ket{1_{m_1},\Psi_{\ell_1\ell_2}^\pm,0_{m_2}} \\
        &= \mp \ket{1_{m_1},\Psi_{\ell_1\ell_2}^\pm,0_{m_2}},\\
    \hat{G}_{m_2} \ket{1_{m_1},\Psi_{\ell_1\ell_2}^\pm,0_{m_2}} &= (\pm 1) (1) \ket{1_{m_1},\Psi_{\ell_1\ell_2}^\pm,0_{m_2}} \\
    &= \pm \ket{1_{m_1},\Psi_{\ell_1\ell_2}^\pm,0_{m_2}},
\end{split}
\end{align}
and similarly for a single excitation in matter site $2$. To understand the dynamics, we consider which of those eigenstates are connected through the Hamiltonian in Eq.~\eqref{eq:plaquette_hamiltonian}. Let us assume first that $h=0$.
We find for the gauge field
\begin{align}
\begin{split}
    \braket{\Phi_{\ell_1\ell_2}^\mp|J(\hat{\sigma}_{\ell_1}^z + \hat{\sigma}_{\ell_2}^z)|\Phi_{\ell_1\ell_2}^\pm} &= 2J,\\
    \braket{\Psi_{\ell_1\ell_2}^\mp|J(\hat{\sigma}_{\ell_1}^z + \hat{\sigma}_{\ell_2}^z)|\Phi_{\ell_1\ell_2}^\pm} &= 0,\\
    \braket{\Psi_{\ell_1\ell_2}^\mp|J(\hat{\sigma}_{\ell_1}^z + \hat{\sigma}_{\ell_2}^z)|\Psi_{\ell_1\ell_2}^\pm} &= 0,\\
    \braket{\Psi_{\ell_1\ell_2}^\pm|J(\hat{\sigma}_{\ell_1}^z + \hat{\sigma}_{\ell_2}^z)|\Psi_{\ell_1\ell_2}^\pm} &= 0.
\end{split}
\end{align}
For the matter field part, we find
\begin{align}
    \braket{0_{m_1},1_{m_2}|\left(\hat{a}_{m_1}\hat{a}_{m_2}^\dagger + {\rm H.c.}\right)|1_{m_1},0_{m_2}} = 1.
\end{align}
This results in tunnelling only for states with Gauge field configuration $\Phi_{\ell_1\ell_2}^\pm$.
i.e., $\ket{1_{m_1},\Phi_{\ell_1\ell_2}^\pm,0_{m_2}} \leftrightarrow \ket{0_{m_1},\Phi_{\ell_1\ell_2}^\mp,1_{m_2}}$ form a two-level system. If we assume $\ket{1_{m_1},\Phi_{\ell_1\ell_2}^+,0_{m_2}}$ as starting state we find 
\begin{equation}
\ket{\varphi(t)} = \cos(2J t) \ket{1_{m_1},\Phi_{\ell_1\ell_2}^+,0_{m_2}} + i\sin(2J t) \ket{0_{m_1},\Phi_{\ell_1\ell_2}^-,1_{m_2}},
\end{equation}
which is Eq.~\eqref{eq:phi_pls_tunneling}.
However, for the gauge field configuration, $\Psi_{\ell_1\ell_2}^\pm$ the tunnelling is forbidden due to Aharonov-Bohm interference.
If we introduce the electric field $|h|>0$, we find the only (additional) non-vanishing term
\begin{align}
    \braket{\Psi_{\ell_1\ell_2}^+|h(\hat{\sigma}_{\ell_1}^x + \hat{\sigma}_{\ell_2}^x)|\Phi_{\ell_1\ell_2}^+} &= 2h.
\end{align}
Thus, under the full Hamiltonian, we find the three-level $\Lambda$-system dynamics
$\ket{1_{m_1},\Phi_{\ell_1\ell_2}^-,0_{m_2}} \stackrel{2J}{\leftrightarrow} \ket{0_{m_1},\Phi_{\ell_1\ell_2}^+,1_{m_2}} \stackrel{2h}{\leftrightarrow} \ket{0_{m_1}, \Psi_{\ell_1\ell_2}^+,1_{m_2}}$, whereas $\ket{\Psi_{\ell_1\ell_2}^-}$ never participates. 
\subsection{Characterising flux and parity of Bell states}\label{sec:supp_charact_flux_and_bell}
\begin{figure}
    \centering
    \includegraphics[width=1\linewidth]{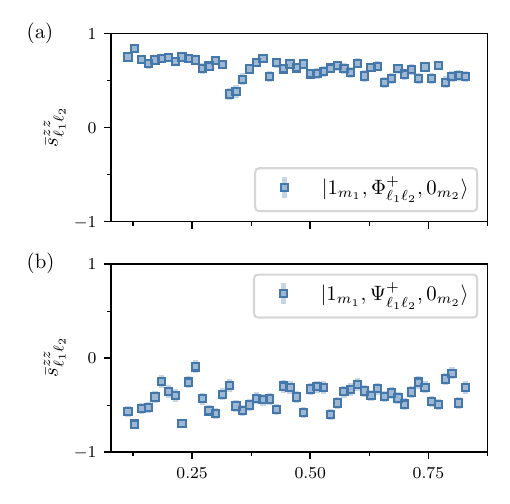}
    \caption{We measure the observable \(\bar{s}_{\ell_1\ell_2}^{zz} = \braket{\hat{\sigma}_{\ell_1}^z \hat{\sigma}_{\ell_2}^z}\), which is related to the flux through the loop, as a function of evolution time. The measurements are performed for the initial states \(\ket{1_{m_1}, \Phi_{\ell_1\ell_2}^+, 0_{m_2}}\) and \(\ket{1_{m_1}, \Psi_{\ell_1\ell_2}^+, 0_{m_2}}\). In the ideal case governed by the loop Hamiltonian \(\hat{H}_{\mathbb{Z}_2}^{\rm loop}\), the flux, and therefore \(\bar{s}_{\ell_1\ell_2}^{zz}\), is expected to remain constant at $1$ for $\ket{\Phi_{\ell_1\ell_2}^+}$ and $-1$ for $\ket{\Psi_{\ell_1\ell_2}^+}$. In our data, \(\bar{s}_{zz}\) stays largely unchanged, consistent with this expectation. However, we observe a slow decay attributed to decoherence, and small kinks in the signal that we attribute to spectral features in the noise present in our system.}
    \label{fig:supp_sz_plaquette}
\end{figure}

Aside from measuring $\bar{s}_{\ell_1\ell_1}^{xx}$ as shown in Fig.~\ref{fig:plaqutte_ab_effect}, we also measure $\bar{s}_{\ell_1\ell_2}^{zz}= \braket{\hat{\sigma}_{\ell_1}^z \hat{\sigma}_{\ell_2}^z}$, which is proportional to the flux through the loop. We show the measurement results in Fig.~\ref{fig:supp_sz_plaquette}. The observable stays constant, apart from decoherence-induced decay.

\begin{figure}
    \centering
    \includegraphics[width=1\linewidth]{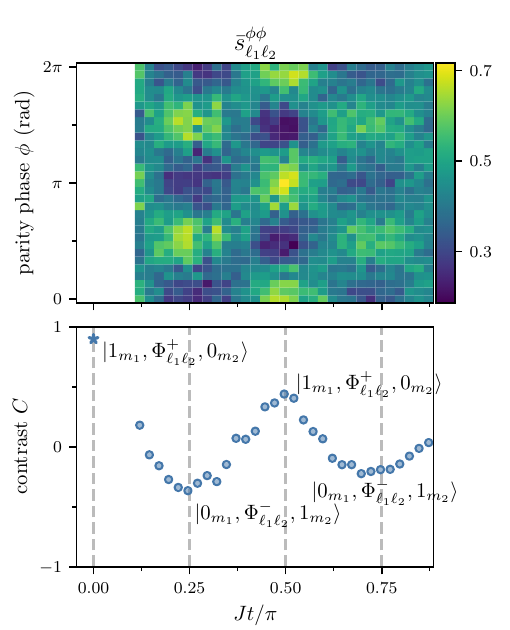}
    \caption{
    Investigating the expectation value $\bar{s}_{\ell_1\ell_2}^{\phi\phi} = \braket{\hat{\sigma}_{\ell_1}^\phi \hat{\sigma}_{\ell_2}^{\phi}}$ as a function of the applied tunnelling duration. In the heatmap, we show the raw data. In the lower plot, we display the extracted contrast found by fitting a sinusoid at every duration. We have indicated, where the measured contrast $C$ is closest to the maximally entangled states of the system. The minimal duration $Jt/\pi=0.24$ is given by the pulse ramping duration. The error bars denoting the 68\% confidence interval in the lower panel are smaller than the marker size. Additionally, in the lower panel, we added the contrast of the initially prepared Bell state $\ket{\Phi_{\ell_1\ell_2}^+}$ (star).
    }
    \label{fig:supp_contrast}
\end{figure}

Further, to understand the gauge-field dynamics precisely for the initial state $\ket{1_{m_1}, \Phi_{\ell_1\ell_2}^+,0_{m_2}}$ 
we can expand the analytic expression by considering Eq.~\ref{eq:phi_pls_tunneling} and tracing out the motion
\begin{align}
    \begin{split}
    \rho &= \cos^2(2J t) \ket{\Phi_{\ell_1\ell_2}^+}\bra{\Phi_{\ell_1\ell_2}^+} +\sin^2(2J t) \ket{\Phi_{\ell_1\ell_2}^-}\bra{\Phi_{\ell_1\ell_2}^-}\\
    &=\frac{1}{2} (\ket{\uparrow_{\ell_1}\uparrow_{\ell_2}}\bra{\uparrow_{\ell_1}\uparrow_{\ell_2}} +\ket{\downarrow_{\ell_1}\downarrow_{\ell_2}}\bra{\downarrow_{\ell_1}\downarrow_{\ell_2}})\\
    &+\frac{\cos(4Jt)}{2}(\ket{\uparrow_{\ell_1}\uparrow_{\ell_2}}\bra{\downarrow_{\ell_1}\downarrow_{\ell_2}}+ \ket{\downarrow_{\ell_1}\downarrow_{\ell_2}}\bra{\uparrow_{\ell_1}\uparrow_{\ell_2}}).
    \end{split}
\end{align}
At $4Jt = \pi/2$ we find the completely disentangled state $\rho = (\ket{\uparrow_{\ell_1}\uparrow_{\ell_2}}\bra{\uparrow_{\ell_1}\uparrow_{\ell_2}} + \ket{\downarrow_{\ell_1}\downarrow_{\ell_2}}\bra{\downarrow_{\ell_1}\downarrow_{\ell_2}})/2$.
We extract the contrast $C = \cos(4Jt)$ as a function of the duration of the tunnelling. To extract the contrast we measure $\bar{s}_{\ell_1\ell_2}^{\phi\phi} = \braket{\cos(\phi) (\hat{\sigma}_{\ell_1}^x \hat{\sigma}_{\ell_2}^x) + \sin(\phi)(\hat{\sigma}_{\ell_1}^y \hat{\sigma}_{\ell_2}^y)}$ and vary $\phi$ (see Fig.~\ref{fig:supp_contrast}). This is also known as a parity measurement. 

As mentioned previously, depending on the technique used to produce the Bell state ${\ket{\Phi_{\ell_1\ell_2}^{+, \tilde\phi}} = \frac{1}{\sqrt{2}} \left( \ket{\uparrow_{\ell_1}, \uparrow_{\ell_2}} + e^{i \tilde{\phi}} \ket{\downarrow_{\ell_1}, \downarrow_{\ell_2}} \right)
}$, a relative phase $\tilde{\phi}$ may arise, which would shift the parity fringe shown in Fig.~\ref{fig:supp_contrast}. It is important to track and account for \(\tilde{\phi}\) to correctly measure the
$\bar{s}_{\ell_1 \ell_2}^{xx}$ correlation, which effectively corresponds to measuring a rotated correlation
$\bar{s}_{\ell_1 \ell_2}^{\tilde\phi \tilde\phi}$.

\section{Proof of $[\hat{H}_{\mathbb{Z}_2}^{\rm link}, \hat{G}_{m_i}] = 0$}
The electric field part $[h\hat{\sigma}_\ell^z, \hat{G}_{m_i}] =0$ can be seen directly. What is left is to prove is $[\hat{\sigma}_\ell^z (\hat{a}_{m_1}\hat{a}_{m_2}^\dagger + \hat{a}_{m_1}^\dagger\hat{a}_{m_2}) , \hat{G}_{m_i}]=0$. 
First let us calculate
\begin{align}
\begin{split}
    ({\rm e}^{{\rm i}\pi \hat{a}^\dagger\hat{a}})^\dagger \hat{a}^\dagger ({\rm e}^{{\rm i}\pi \hat{a}^\dagger\hat{a}})\ket{n} &= ({\rm e}^{{\rm i}\pi \hat{a}^\dagger\hat{a}})^\dagger \hat{a}^\dagger (-1)^n \ket{n} \\
    &= ({\rm e}^{{\rm i}\pi \hat{a}^\dagger\hat{a}})^\dagger (-1)^n  \sqrt{n+1}\ket{n+1} \\
    &= (-1)^{n+1} (-1)^n \sqrt{n+1}\ket{n+1}\\
    &= -\hat{a}^\dagger \ket{n},
\end{split}
\end{align}
and similarly
\begin{align}
        (e^{i\pi \hat{a}^\dagger\hat{a}})^\dagger \hat{a} (e^{i\pi \hat{a}^\dagger\hat{a}})\ket{n}= -\hat{a} \ket{n}.
\end{align}
Thus, we can make use of the identities
\begin{align}
\hat{\sigma}_\ell^x \hat{\sigma}_\ell^z\hat{\sigma}_\ell^x &= - \hat{\sigma}_\ell^z  \label{eq:sigma_xz}\\
    ({\rm e}^{{\rm i}\pi \hat{a}_{m_1}^\dagger\hat{a}_{m_1}})^\dagger (\hat{a}_{m_1}\hat{a}_{m_2}^\dagger + \hat{a}_{m_1}^\dagger\hat{a}_{m_2}){\rm e}^{{\rm i}\pi \hat{a}_{m_1}^\dagger\hat{a}_{m_1}} &= - (\hat{a}_{m_1}\hat{a}_{m_2}^\dagger+\hat{a}_{m_1}^\dagger\hat{a}_{m_2}), \label{eq:ad_parity}  
\end{align}
to prove that for the generator $\hat{G}_{m_1}$
\begin{align}
    &\underbrace{(\hat{\sigma}_{\ell}^x{\rm e}^{{\rm i}\pi \hat{a}_{m_1}^\dagger\hat{a}_{m_1}})^\dagger}_{\hat{G}_{m_1}^\dagger} \hat{H}_{\mathbb{Z}_2}^{\rm link} \underbrace{(\hat{\sigma}_\ell^x{\rm e}^{{\rm i}\pi \hat{a}_{m_1}^\dagger\hat{a}_{m_1}})}_{\hat{G}_{m_1}} = \\&=  {\rm e}^{-{\rm i}\pi \hat{a}_{m_1}^\dagger\hat{a}_{m_1}} \hat{\sigma}_\ell^x \hat{H}_{\mathbb{Z}_2}^{\rm link} \hat{\sigma}_\ell^x{\rm e}^{{\rm i}\pi \hat{a}_{m_1}^\dagger\hat{a}_{m_1}}\\
    &\stackrel{\ref{eq:sigma_xz}}{=} - {\rm e}^{-{\rm i}\pi \hat{a}_{m_1}^\dagger\hat{a}_{m_1}}  \hat{H}_{\mathbb{Z}_2}^{\rm link} {\rm e}^{{\rm i}\pi \hat{a}_{m_1}^\dagger\hat{a}_{m_1}}\\
    &\stackrel{\ref{eq:ad_parity}}{=} \hat{H}_{\mathbb{Z}_2}^{\rm link},
\end{align}
and the same for $\hat{G}_{m_2}$. This extends to $[\hat{H}_{\mathbb{Z}_2}^{\rm loop}, \hat{G}_{m_i}]=0$ as well.

\section{Additional investigation and Future work}
\subsection{Competition between electric field, tunnelling, and Aharonov-Bohm effect}
\begin{figure}
    \centering
    \includegraphics[width=1\linewidth]{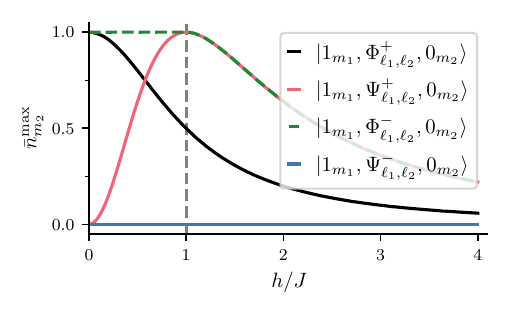}
    \caption{Numerical simulation of maximal tunnelling magnitude  $\bar{n}_{m_2}^{\rm max} = \max(\braket{n_{m_2}})$ for different ratios of $h/J$ for the four possible initial entangled gauge field configurations of the loop. In the limit of $h \rightarrow \infty$ the tunnelling magnitude vanishes for all the starting states. For $h = 0$, the tunnelling magnitude is maximised for the Bell-states $\Phi_{\ell_1\ell_2}^\pm$ and completely suppressed for $\Psi_{\ell_1\ell_2}^\pm$ as a result of the AB-interference. For $\Psi_{\ell_1\ell_2}^+$, however, any finite electric field perturbs the perfect destructive AB-interference, and tunnelling occurs.}
    \label{fig:supp_3effect_competition}
\end{figure}

As in the case of a single ion, there is an inherent competition between the electric field term and the tunnelling term. However, there is an additional dynamic on the loop due to the Aharonov-Bohm interference. These three effects now all compete, leading to non-trivial dynamics.
We briefly investigate the full $\Lambda$-system dynamics $\ket{1_{m_1},\Phi_{\ell_1\ell_2}^-,0_{m_2}} \stackrel{2J}{\leftrightarrow} \ket{0_{m_1},\Phi_{\ell_1\ell_2}^+,1_{m_2}} \stackrel{2h}{\leftrightarrow} \ket{0_{m_1},\Psi_{\ell_1\ell_2}^+,1_{m_2}}$ and the isolated $\ket{\Psi_{\ell_1\ell_2}^-}$ state using numerical simulations. The effect of a non-vanishing electric field on the dynamics is shown in Fig.~\ref{fig:supp_3effect_competition}. 

We calculate the maximal tunnelling amplitude $\bar{n}_{m_2,{\rm max}}$ for different ratios of $h/J$ considering starting state $\ket{ 1_{m_1}, \Phi_{\ell_1\ell_2}^\pm,0_{m_2}}$ and $\ket{1_{m_1},\Psi_{\ell_1\ell_2}^\pm,0_{m_2}}$. For $\ket{1_{m_1},\Phi_{\ell_1\ell_2}^+,0_{m_2}}$, where both gauge fields are aligned, we find the same dynamics as on a single gauge link; the tunnelling magnitude is gradually suppressed as $h\rightarrow \infty$. For $\ket{1_{m_1},\Phi_{\ell_1\ell_2}^-,0_{m_2}}$ the tunnelling magnitude is maximum at $h=0$ and stays maximum until $h>J$ after which it is gradually suppressed as $h\rightarrow \infty$.
However, for the case where the gauge fields are anti-aligned ($\ket{\Psi_{\ell_1\ell_2}^\pm}$) and hence $\phi_{AB}= \pi$, the tunnelling is fully suppressed for $h=0$ as well as $h\rightarrow \infty$. For $\ket{\Psi_{\ell_1\ell_2}^-}$, the destructive interference is perfect for any $h$. But, for $\ket{\Psi_{\ell_1\ell_2}^+}$, there is an intermediate regime (centred at $J=h$) where tunnelling is possible. Here, the presence of the electric field term destroys the perfect destructive interference due to the AB effect. 

\subsection{Even and odd parity squeezed matter}\label{sec:supp_squeezed}
\begin{figure}
    \centering
    \includegraphics[width=1\linewidth]{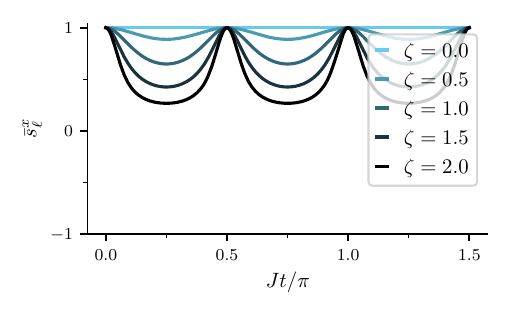}
    \caption{Expectation value of $\bar{s}_\ell^x$ as a function of tunnelling duration, starting from the state $\ket{\zeta_{m_1}^+, +_\ell, 0_{m_2}}$, representing no  matter excitation at either site, i.e, \( q_{m_1} = q_{m_2}=  0 \), with variable squeezing $\zeta$ applied to $\ket{0_{m_1}}$. The resulting correlation changes from no tunnelling at all to a pattern of partial tunnelling with full revival at each swap duration.}
    \label{fig:supp_sim_squeezed_00_plus}
\end{figure}

\begin{figure}
    \centering
    \includegraphics[width=1\linewidth]{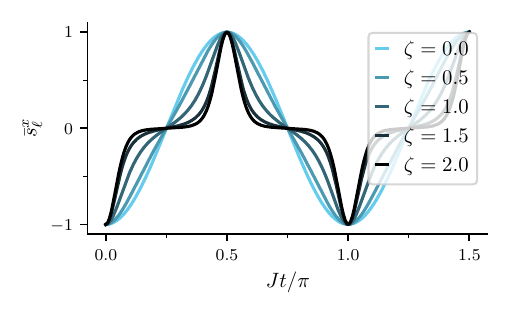}
    \caption{Expectation value of $\bar{s}_\ell^x$ as a function of tunnelling duration, starting from the state $\ket{\zeta_{m_1}^-, -_\ell, 0_{m_2}}$, representing a matter excitation at $m_1$, i.e., \( q_{m_1} = 1, q_{m_2}=0 \), with variable squeezing $\zeta$ applied to $\ket{1_{m_1}}$. The resulting correlation changes from sinusoidal to an increasingly sharp feature which in the (nonphysical) limit of infinite squeezing $\zeta \rightarrow \infty$ becomes a delta function at the full interchange times.}
    \label{fig:supp_sim_squeezed_10_plus}
\end{figure}

A strength of our mapping is that the matter sites are true bosons. As such, they can support more than a single excitation. 
For the $\mathbb{Z}_2$ LGT 
we are still interested in states that have a well-defined charge and hence are eigenstates of $\hat{P}_{m_i} \ket{\zeta_{m_i}^\pm} \equiv \exp({\rm i} \pi \hat{a}_{m_i}^\dagger \hat{a}_{m_i}) \ket{\zeta_{m_i}^\pm} = \pm \ket{\zeta_{m_i}^\pm}$. A particular subset of such states are the squeezed vacuum and squeezed phonon-added vacuum states~\cite{liu2000photonadded}, namely
\begin{align}
    \begin{split}
    \ket{\zeta_{m_i}^+} &= \hat{S}(\zeta)\ket{0_{m_i}} \\
    &= \frac{1}{\sqrt{\cosh(r)}} \sum_{n=0}^{\infty} \frac{\sqrt{(2n)!}}{2^n n!} ({\rm e}^{{\rm i}\theta} \tanh(r))^n \ket{2n_{m_i}},
    \end{split}\\
    \begin{split}
    \ket{\zeta_{m_i}^-} &= \hat{S}(\zeta)\ket{1_{m_i}} \\
    &= \frac{1}{({\cosh(r)})^{3/2}} \sum_{n=0}^{\infty}\frac{\sqrt{(2n +1)!}}{2^n n!} ({\rm e}^{{\rm i}\theta} \tanh(r))^n \ket{2n+1_{m_i}},
    \end{split}
\end{align}
where the squeezing operator is defined as
\begin{align}
    \hat{S}(\zeta) &= \exp\left(\frac{1}{2} \big(\zeta^\ast\hat{a}_{m_i}^2 - \zeta(\hat{a}^\dagger_{m_i})^2\big)\right),
\end{align}
with squeezing parameter $\zeta$.
These states are eigenstates of the gauge-symmetry operator, and can thus be associated with dynamical charges $q_{m_i}=0$ and $q_{m_i}=1$, respectively. As in the single excitation case, it is of interest to study the expectation values of the observables $\bar{n}_{m_1}$, $\bar{n}_{m_2}$ and $\bar{s}_\ell^x$. In particular, we consider the initial state $\ket{ \zeta_{m_1}^{\pm},\pm_{\ell}, 0_{m_2}}$ and study the tunnelling (i.e., no transverse field term). While the mean phonon number $\bar{n}_{m_1}$, $\bar{n}_{m_2}$ number oscillates back and forth in a sinusoidal shape between the two sites, the dynamics of the qubit correlation changes as a function of $|\zeta|$ (see Fig.~\ref{fig:supp_sim_squeezed_00_plus}, \ref{fig:supp_sim_squeezed_10_plus}).


\begin{figure}
    \centering
    \includegraphics[width=1\linewidth]{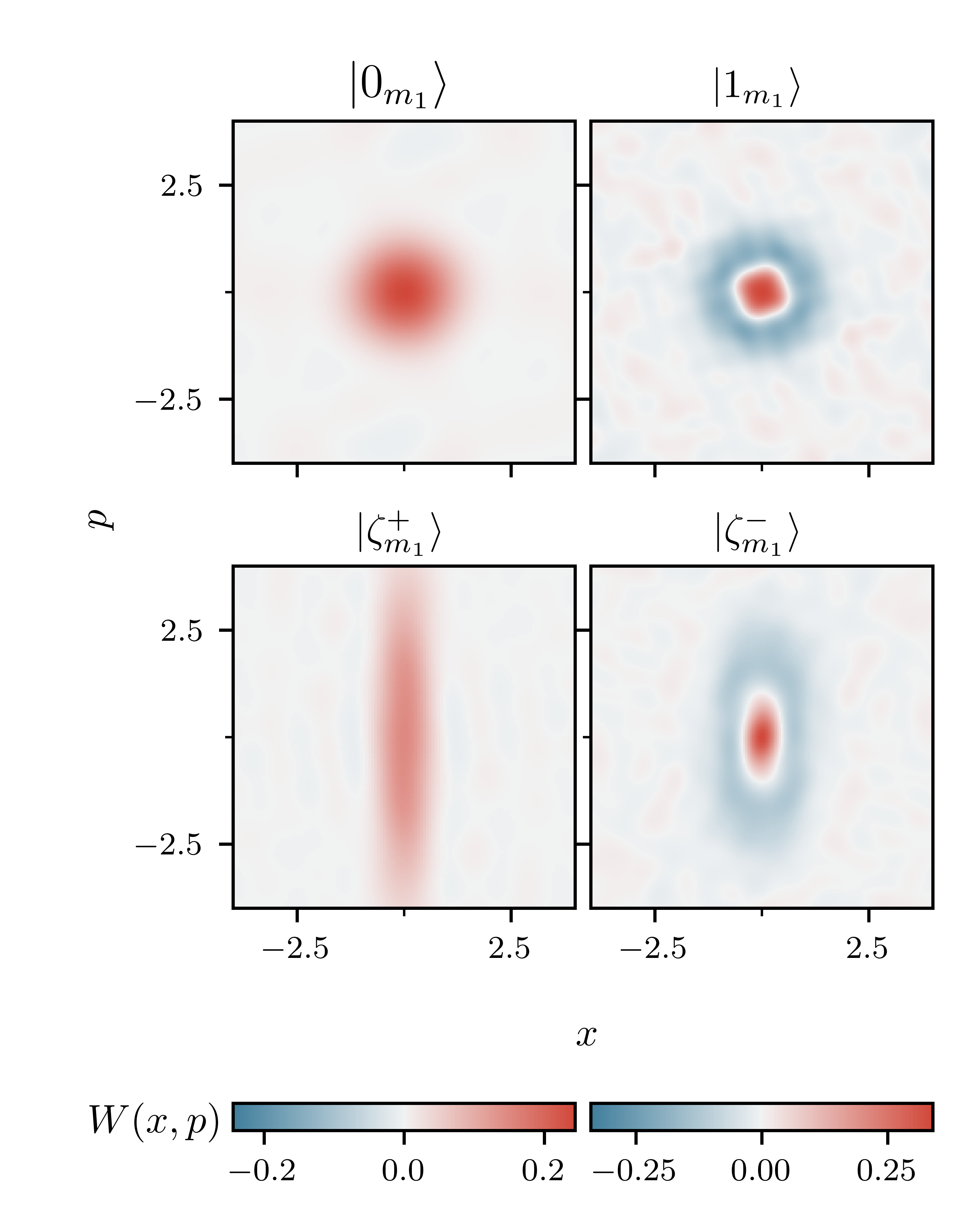}
    \caption{Reconstructed Wigner function for a matter site in state $\ket{0_{m_1}}$, $\ket{1_{m_1}}$,  $\hat{S}(\zeta)\ket{0_{m_1}} =\ket{\zeta_{m_1}^{+}}$ with $\zeta \approx 1.1$, and $\hat{S}(\zeta)\ket{1} = \ket{\zeta_{m_1}^{-}}$ with $\zeta=0.96(2)$. The reconstruction of $\ket{1_{m_1}}$ suffers from undersampling. As a result, the shape is distorted by the sampling grid pattern (square).}
    \label{fig:supp_wigner_sqz_fock}
\end{figure}
In a first instance we verify that we can synthesise these states in the experiment. In Fig.~\ref{fig:supp_wigner_sqz_fock} we display the reconstructed Wigner function of $\ket{\zeta_{m_1}^{\pm}}$. After this, we verify that the squeezed matter tunnels as we expect. We prepare a state $\ket{\zeta_{m_1}^{+},\downarrow_{\ell}, 0_{m_2}}$ which is initially squeezed on matter site 1 while prepared in its vacuum state on matter site $m_2$. 
We measure the squeezing magnitude $|\zeta|$ (using the method described in Refs.~\cite{lo2015spin, bazavan2024easy, saner2025quantum}) on each matter site as a function of the applied tunnelling duration. We find that as matter site $m_1$, which is initially squeezed, gradually becomes less squeezed, the degree of squeezing increases on matter site $m_2$. Note, that the intermediate states at site $m_1$ are a superposition between the squeezed and the vacuum state (i.e., $a\ket{r}+ b\ket{0}$). Only, for $a=0$ or $b=0$ is this state is a true squeezed state with well-defined squeezing magnitude. Nonetheless, we use it as a proxy to quantify the changing state. Subsequently, we reconstruct the full Wigner function of each mode at different instances of the tunnelling duration using the technique described in Ref.~\cite{fluhmann2020direct}.

\begin{figure}
    \centering
    \includegraphics[width=1\linewidth]{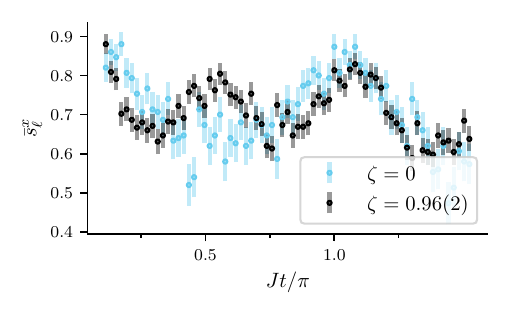}
    \caption{Experimentally measured expectation value of \( \bar{s}_\ell^x \) as a function of tunnelling duration, starting from the state \( \ket{\zeta_{m_1}^+, +_\ell, 0_{m_2}} \) with squeezing \( \zeta \) applied to \( \ket{0_{m_1}} \), representing no  matter excitation at either site, i.e, \( q_{m_1} = q_{m_2}=  0 \). For \( \zeta = 0 \) (blue points), this corresponds to a thermal state near the ground state \( \ket{0_{m_i}} \). For \( |\zeta| = 0.96(2) \) (black points), the matter field is squeezed. In the unsqueezed case, we observe periodic spin correlation with $J t/\pi = 1$, corresponding to regular tunnelling of the thermal states. When squeezing is applied, a partial revival is observed at $J t/\pi = 0.5$, as predicted by simulations. For easier comparison, we adjusted the baseline of the squeezed population dynamics. The error bars denote the 68\% confidence interval. 
        }
    \label{fig:supp_plus_sqz00_sx_exp}
\end{figure}

\begin{figure}
    \centering
    \includegraphics[width=1\linewidth]{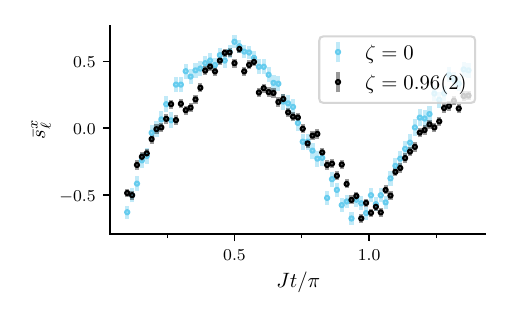}
    \caption{Experimentally measured expectation value of \( \bar{s}_\ell^x \) as a function of tunnelling duration, starting from the state \( \ket{\zeta_{m_1}^-, -_\ell, 0_{m_2}} \) with squeezing \( \zeta \) applied to \( \ket{0_{m_1}} \), representing a matter excitation at $m_1$, i.e., \( q_{m_1} = 1, q_{m_2}=0 \). For \( \zeta = 0 \) (blue points), this corresponds to a thermal state near the ground state \( \ket{0_{m_i}} \). For \( |\zeta| = 0.96(2) \) (black points), the matter field is squeezed.
    In the unsqueezed case, we observe spin periodicity of $J t/\pi = 1$. With squeezing, the feature sharpens slightly, though less pronounced than predicted. To highlight the deviation from a sinusoidal pattern, the contrast of the squeezed dynamics was scaled by a factor of 1.45. The reduced contrast suggests decoherence affects the dynamics. The error bars denote the 68\% confidence interval.     
    }
    \label{fig:supp_plus_sqz10_sx_exp}
\end{figure}

We proceed to measure the gauge field  correlation of the tunnelling for the initial configurations $\ket{\zeta_{m_1}^{+}, +_{\ell},0_{m_2}}$ (Fig.~\ref{fig:supp_plus_sqz00_sx_exp}) and $\ket{\zeta_{m_1}^{-}, -_{\ell}, 0_{m_2}}$ (Fig.~\ref{fig:supp_plus_sqz10_sx_exp}). While we do find that the pattern of the gauge field correlation resembles the simulated prediction it is substantially less pronounced. We suspect that this is a result of the finite motional and spin coherence times in our system. Investigating this problem from a numerical side is surprisingly challenging. In particular, we find that we need to choose a Fock state truncation of at least $25$ to faithfully represent the squeezed state. This gives us a resulting hybrid state size of $2 \times 25\times 25 = 1250$ which can rapidly become challenging to simulate, particularly if we consider larger lattices.

\begin{figure}
    \centering
    \includegraphics[width=1\linewidth]{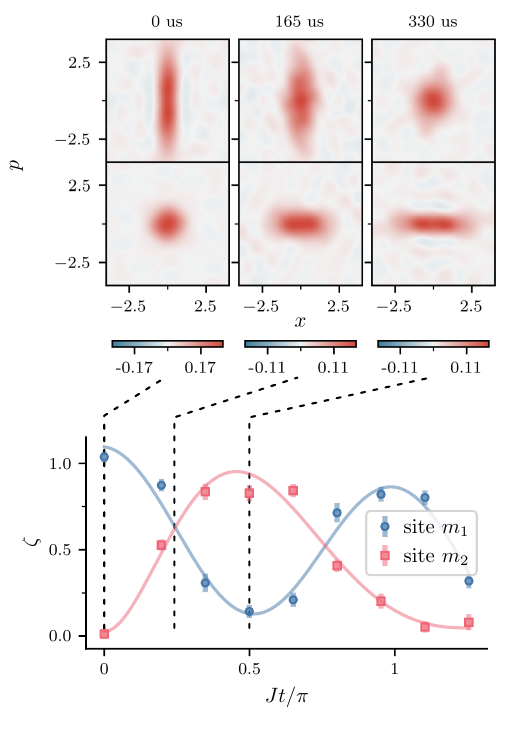}
    \caption{Observed tunnelling from the initial state $\ket{\zeta_{m_1}^+,\downarrow_\ell,0_{m_2}}$. We extract the squeezing magnitude $\zeta$ at each matter site $m_i$ as a function of tunnelling duration using the method of Ref.~\cite{lo2015spin}; error bars indicate uncertainties from the reconstruction. We also show decaying sinusoidal fits (solid lines). At selected durations, we additionally perform full Wigner function reconstructions using the technique of Ref.~\cite{fluhmann2020direct}.   
}
    \label{fig:supp_measure_sqz_mag_vs_duration}
\end{figure}

\clearpage

\label{suppl_material}
\end{document}